\documentclass[12pt]{article}
\usepackage[colon]{natbib}
\usepackage[pdftex,colorlinks=true]{hyperref}
\usepackage{verbatim}  
\hypersetup{citecolor=blue,linkcolor=blue,urlcolor=blue}
\usepackage{setspace}

\usepackage{rotating,multirow,multicol,bm}
\usepackage{lscape,pdflscape}
\usepackage{algorithm,algpseudocode}
\usepackage{lipsum}
\usepackage[margin=0.85in]{geometry}
\usepackage{setspace}
\graphicspath{{Figures/}}
\usepackage{graphicx}
\usepackage{caption, subcaption}
\usepackage{booktabs}
\usepackage{longtable}
\usepackage[title]{appendix}
\usepackage{makecell}
\usepackage{threeparttable} 
\usepackage{algorithm}
\usepackage{chngcntr} %

\usepackage{amsmath,paralist,color,url}
\usepackage[charter]{mathdesign}

\newtheorem{theorem}{Theorem}
\newtheorem{lemma}{Lemma}
\newtheorem{proposition}{Proposition}

\newtheorem{corollary}{Corollary}

\DeclareCaptionFormat{cont}{#1 (cont.)#2#3\par}

\title{\vspace{-1.2cm} }
\author{}
\date{ }

\begin{document}

\renewcommand{\thefootnote}{*}

\begin{center}
{\Large Gaussian Process Stochastic Frontiers with Two-Way Fixed Effects}

\vspace{0.5cm}
{\large Puguang Nie}\textsuperscript{~a}, \hspace{0.0cm}
{\large Jianxin Pan}\textsuperscript{~a}, \hspace{0.0cm}  
{\large Xibin Zhang}\textsuperscript{~b}\footnote{Address: 900 Dandenong Road, Caulfield East, Victoria 3145, Australia. Telephone: +61 3 99032130. \\ Email: {\em\color{blue}xibin.zhang@monash.edu}.}

\vspace{0.5cm}

$^\text{a}$ Guangdong Provincial Key Laboratory of Interdisciplinary Research and Application for Data Science, Beijing Normal-Hong Kong Baptist University, Zhuhai, 519087, China

\vspace{0.1in}

$^\text{b}$ Department of Econometrics and Business Statistics, Monash University, Australia

\vspace{0.5cm}

November 24, 2026 
\end{center}

\vspace{0.3cm}
\baselineskip 0.78cm

\noindent{\bf Abstract}: Parametric frontier restrictions and unobserved time heterogeneity can distort efficiency measurement in panel data. This paper develops a Bayesian stochastic frontier model with a Gaussian process prior and unrestricted individual and time effects that may be correlated with inputs. Conditional on inefficiency, orthogonal projection eliminates both sets of fixed effects, yielding a restricted likelihood that retains observation-specific inefficiency. Monte Carlo experiments show competitive performance under correct parametric specification and improved performance when benchmark models impose an incorrect frontier functional form or omit time effects. An application to Spanish dairy farms observed from 1999 to 2010 yields similar average technical efficiency across specifications, but a lower posterior mean of average returns to scale, with greater uncertainty, under the Gaussian process frontier. Models with year effects also yield more stable estimated efficiency profiles than the benchmark with a quadratic trend and time–input interactions. These findings show that conclusions about scale economies and efficiency dynamics depend on both frontier flexibility and the treatment of time heterogeneity.

\medskip 



\vspace{4cm}

\renewcommand{\thefootnote}{\arabic{footnote}}
\setcounter{footnote}{0}

\pagebreak 
\baselineskip 0.78cm 

\section{Introduction}
\label{sec:introduction}

Stochastic frontier analysis (SFA) estimates production technology and
technical efficiency while allowing for statistical noise. Since
\citet{Aigner1977} and \citet{Meeusen77}, the literature has developed
increasingly flexible models for cross-sectional and panel data
\citep[see][for reviews]{Kumbhakar2022a,Kumbhakar2022b}.
Efficiency measurement nevertheless depends critically on the production
function and the treatment of unobserved heterogeneity. Restrictions on
either component can cause variation in technology or production
conditions to be attributed to inefficiency.

Conventional models specify a Cobb--Douglas, translog, or another
finite-dimensional production function
\citep{Battese1992,Battese1995,George1997}. Although a translog frontier
allows more curvature than a Cobb--Douglas specification, it remains a
global approximation that restricts how elasticities vary across inputs.
Efficiency estimates can be sensitive to these restrictions
\citep{giannakas2003}. This concern has motivated semiparametric and
nonparametric frontier methods
\citep{fan1996,KUMBHAKAR2007,SUN2013305,FERRARA2017761,Yao2019,
wang_flexible_2024,parmeter2025}. Bayesian contributions include the
Gaussian process frontier of \citet{Emvalomatis2020} and the penalized
spline approach of \citet{klein_modelling_2020}.

Panel data introduce a second specification problem. Persistent differences
in location, production environments, management, or technology should
not automatically be classified as time-varying inefficiency. Fixed
effects allow for producer-specific heterogeneity without assigning it
a probability distribution or imposing independence from inputs.
The true fixed-effects model of \citet{GREENE2005} and subsequent work
\citep{WANG2010286,CHEN201465,BELOTTI2018,bellio_2024} emphasize this
distinction. Additive time effects can similarly accommodate annual
shifts shared by producers that a smooth deterministic trend may miss.

Recent work combines flexible frontiers with panel heterogeneity.
\citet{wang_flexible_2024} use first differences to remove individual
effects and estimate an additive frontier with bivariate interactions.
\citet{parmeter2025} study nonparametric identification and estimation
of a four-component model through differencing. These approaches focus
on conditional mean functions for inefficiency. Our focus is the joint
treatment of a flexible common frontier and unrestricted individual and
time effects while retaining a distributional model for transient
inefficiency and posterior inference for observation-specific efficiency.

We develop a Bayesian Gaussian process (GP) stochastic frontier with
two-way fixed effects. A GP prior regularizes the unknown production
function directly \citep{Hagan1978,Rasmussen2005,van2008}, with separate
length scales controlling variation along each input dimension.
Related economic applications use GP priors for nonlinear relationships
in policy evaluation and macroeconomic dynamics
\citep{KASY2018,hauzenberger2025}. The prior provides joint uncertainty
statements for frontier values and derivatives, but does not itself
resolve confounding between the frontier and fixed effects. Structural
identification must therefore be established separately from Bayesian
regularization.

Combining a GP frontier, latent inefficiency, and unrestricted fixed
effects also creates a computational problem. Direct sampling adds
$N+T-2$ fixed-effect coefficients to the latent frontier and inefficiency
vectors and can induce strong posterior dependence. We instead project
the data onto a complete set of orthogonal contrasts that removes both
sets of effects. The frontier retains its interpretation in the original
input space. The resulting likelihood is a restricted likelihood
\citep{PattersonThompson1971} and is proportional to the likelihood
integrated over the fixed effects under Lebesgue measure
\citep{Harville1974,Berger1999}. It uses the residual dimension of the
contrast space, whereas ordinary profiling retains the full sample
dimension in the Gaussian normalization.

The paper makes three contributions. First, it combines a GP production
frontier with unrestricted individual and time effects that may be
related to the inputs. Under explicit conditions on the frontier class
and joint input distribution, we establish identification of the common
frontier and the two error scales. With differentiability and interior
support, the frontier derivatives are also identified. These conditions
distinguish structural identification from the prior regularization
used to recover frontier values at the observed design.

Second, conditional on inefficiency, we eliminate the fixed-effect
coefficients exactly and characterize the resulting covariance scores.
Under the GP-integrated Gaussian model, the eliminated score has mean
zero, while the ordinary profile score has an explicit adjustment term.
The noise-variance update uses the residual degrees of freedom.
These results clarify the role of elimination in the incidental-parameter
problem \citep{LANCASTER2000}; they do not, by themselves, establish
fixed-$T$ consistency for an estimator that also learns the frontier and
latent inefficiencies. The construction extends the use of contrasts
in parametric individual-effects frontiers \citep{CHEN201465} to a GP
frontier with two-way effects.

Third, we develop Bayesian inference that retains inefficiency in the
original observation space. The original inefficiencies have independent
one-sided priors, whereas their projected coordinates need not be positive
or independent. A partially collapsed Gibbs sampler updates the kernel
parameters with the frontier integrated out and then restores its
conditional Gaussian distribution. Joint posterior derivative inference
provides input elasticities and returns to scale, including covariance
across inputs and evaluation points. Eliminating the effects reduces the number of parameters sampled,
although dense GP covariance calculations remain.

Monte Carlo experiments show a modest cost of frontier flexibility when
the translog specification is correct and substantial gains when benchmark
models impose an incorrect frontier or omit time effects. An application
to 90 dairy farms in Asturias observed over 1999--2010 yields similar
average efficiency across specifications, but a lower posterior mean of
average returns to scale, with greater uncertainty, under the GP frontier.
Models with year effects also yield more stable estimated efficiency
profiles than the trend benchmark. These findings illustrate sensitivity
to frontier shape and temporal specification; the empirical comparisons
do not identify the shocks driving these differences.

The remainder of the paper is organized as follows.
Section~\ref{sec:model} presents the model, the identification conditions,
and the properties of the invariant likelihood.
Section~\ref{sec:bayes} develops posterior inference.
Sections~\ref{sec:simulation} and~\ref{sec:application} report the
simulation study and the empirical application, respectively.
Section~\ref{sec:conclude} concludes.
Appendix~\ref{app:section2_proofs} gives the proofs of the theoretical
results, Appendix~\ref{app:computational_details} provides implementation
details, and Appendix~\ref{app:translog} presents additional details
of the empirical study. 

\section{Gaussian process frontier with two-way fixed effects}
\label{sec:model}

This section separates three issues: identification of the common production
frontier, elimination of the two sets of fixed effects, and Gaussian process regularization of the frontier. Structural identification concerns a fixed unknown function. The covariance-score results concern the Gaussian model obtained by integrating that function under its GP prior. These are different probability models, and their conclusions are stated separately.
All proofs are collected in Appendix~\ref{app:section2_proofs}.

\subsection{Panel stochastic frontier model}
\label{subsec:model}

Consider a balanced panel of $N\geq2$ production units observed over
$T\geq2$ periods. Let $y_{it}$ be log output and let
$x_{it}\in\mathcal X\subseteq\mathbb R^p$ collect production inputs or their
specified transformations. The model is
\begin{equation*}
y_{it}=\alpha_i+\gamma_t+f(x_{it})-u_{it}+v_{it},
\qquad i=1,\ldots,N,\quad t=1,\ldots,T.
\end{equation*}
Here $f$ is the common structural frontier, $\alpha_i$ and $\gamma_t$ are
individual and time effects, $u_{it}\geq0$ is transient technical
inefficiency, and $v_{it}$ is statistical noise. Conditional on the complete
input array $X$, the error components satisfy
\begin{equation}
\label{eq:u_dist}
v_{it}\stackrel{\mathrm{iid}}{\sim}N(0,\sigma_v^2),
\qquad
u_{it}\stackrel{\mathrm{iid}}{\sim}N^+(0,\sigma_u^2),
\qquad \sigma_v^2>0,\quad \sigma_u^2\geq0,
\end{equation}
and the two arrays are independent. Their conditional distributions do not
depend on $X$. The notation $N^+(0,\sigma_u^2)$ denotes the distribution of
$\sigma_u|Z|$ for $Z\sim N(0,1)$; at $\sigma_u^2=0$ it is a point mass at
zero. Thus $\sigma_u^2$ is the squared half-normal scale, whereas
$\operatorname{Var}(u_{it})=(1-2/\pi)\sigma_u^2$.

No distribution is imposed on the fixed effects, and no independence
restriction between them and the inputs is used. Impose
\begin{equation*}
\sum_{i=1}^N\alpha_i=0,
\qquad \sum_{t=1}^T\gamma_t=0.
\end{equation*}
The level of $f$ therefore contains the common intercept. These
normalizations separate the frontier level from the two sets of effects;
they do not remove the need for an identification condition on $f$.

\subsection{Matrix representation and orthogonal contrasts}
\label{subsec:matrix}

Let $n=NT$ and stack observations unit by unit. Choose orthonormal contrast
matrices $C_N\in\mathbb R^{N\times(N-1)}$ and
$C_T\in\mathbb R^{T\times(T-1)}$ such that
\[
C_N'C_N=I_{N-1},\quad C_N'\mathbf1_N=0,
\qquad C_T'C_T=I_{T-1},\quad C_T'\mathbf1_T=0.
\]
Write $\alpha=C_Na$, $\gamma=C_Tc$, and define $D=[C_N\otimes\mathbf1_T,\ \mathbf1_N\otimes C_T]$ and $\delta=(a',c')'$. We then have 
\begin{equation}
\label{eq:DTD}
D'D=\operatorname{diag}(TI_{N-1},NI_{T-1}),~~ \text{ and }~~
D'\mathbf1_n=0.
\end{equation}
With $f_X=[f(x_{11}),\ldots,f(x_{NT})]'$, the stacked model is
\begin{equation*}
y=D\delta+f_X-u+v.
\end{equation*}
Let $\mathcal C(D)$ be the column space of $D$ and define
\begin{equation*}
M_D=I_n-D(D'D)^{-1}D'.
\end{equation*}
For the balanced panel, let the number of free fixed-effect coefficients be $q=N+T-2$ and set
\[
q_0=n^{-1/2}\mathbf1_n,\qquad Q=C_N\otimes C_T,
\qquad R=[q_0,Q],\qquad r=n-q=(N-1)(T-1)+1,
\]
where $r$ is the dimension remaining after eliminating the fixed effects. Thus,
\[
R\in\mathbb R^{n\times r},
\qquad R'y\in\mathbb R^r.
\]
This dimension determines the Gaussian likelihood's normalizing factor $(2\pi\sigma_v^2)^{-r/2}$ and the noise-variance update's shape parameter $a_v+r/2$.

\begin{proposition}[Geometry of fixed-effect elimination]
\label{prop:contrast_geometry}
The matrix $D$ has rank $q$, and
\begin{equation}
\label{eq:Rproperties}
R'R=I_r,\qquad R'D=0,\qquad RR'=M_D.
\end{equation}
Moreover, $R'y$ is a maximal invariant under the translations
$y\mapsto y+Db$, $b\in\mathbb R^q$: two data vectors have the same
contrasts if and only if they differ by an element of $\mathcal C(D)$.
The grand-mean direction $q_0$ is retained because the common intercept
belongs to $f$, not to $D$.
\end{proposition}

\noindent{\em Proof}. See Appendix~\ref{app:proof_contrast_geometry}. 

The subsequent matrix and likelihood results apply to any fixed,
full-column-rank $n\times q$ nuisance design with $q<n$ and any orthonormal
basis $R$ for $\mathcal C(D)^\perp$. For an unbalanced panel, the design and
its normalizations must be specified on the observed cells; QR
decomposition then constructs $R$. The explicit Kronecker formula above
is specific to a balanced panel.

\subsection{Structural identification}
\label{subsec:identification}

Fixed-effect elimination preserves the part of a frontier that varies
outside the nuisance space. It cannot distinguish two admissible
frontiers whose evaluations differ only by fixed effects. Let
$\mathcal F$ be the admissible class, let
$\mathcal F-\mathcal F=\{f_1-f_2:f_1,f_2\in\mathcal F\}$, and define
$\mu_X=n^{-1}\sum_{i,t}\mathcal L(X_{it})$, the mixture of cell input laws.

\medskip
\noindent\textbf{Assumption 1 (No functional confounding).}
For every $h\in\mathcal F-\mathcal F$,
\begin{equation*}
M_Dh_X=0\ \text{almost surely}
\quad\Longrightarrow\quad
h=0\ \ \mu_X\text{-almost everywhere},
\end{equation*}
where $h_X=[h(X_{11}),\ldots,h(X_{NT})]'$.

The conclusion is equivalent to $h_X=0$ almost surely. Assumption~1 is a
population restriction on the common function and the joint input law;
it is not a rank claim about unrestricted values assigned at a single
realized design. In particular, it is stronger than excluding a
representation with fixed, nonrandom cell coefficients.

For a primitive sufficient condition, let
$\mathcal S_{it}=\operatorname{supp}(X_{it})$ be the topological cell
support and let $\mathcal S=\bigcup_{i,t}\mathcal S_{it}$. Construct a graph
on the panel cells with an edge whenever the two cell supports intersect.
A rectangle consists of four distinct cells
$(i,t),(i,s),(j,t),(j,s)$ with $i\ne j$ and $t\ne s$.

\begin{proposition}[Primitive sufficient condition]
\label{prop:primitive_identification}
Suppose every $h\in\mathcal F-\mathcal F$ is continuous on $\mathcal S$
and the support-overlap graph is connected. Suppose there is a collection
of rectangles covering all panel cells such that, for every selected
rectangle,
\begin{equation}
\label{eq:rectangle_support}
\operatorname{supp}(X_{it},X_{is},X_{jt},X_{js})
=\mathcal S_{it}\times\mathcal S_{is}\times
  \mathcal S_{jt}\times\mathcal S_{js}.
\end{equation}
Then Assumption~1 holds. In fact, $M_Dh_X=0$ almost surely implies
$h(x)=0$ at every $x\in\mathcal S$.
\end{proposition}

\noindent{\em Proof}. See Appendix~\ref{app:proof_primitive_identification}.

Condition~\eqref{eq:rectangle_support} allows dependence but excludes
support restrictions that make one cell's input a deterministic function
of the other cells in the rectangle. It is sufficient, not necessary.
It lets the rectangle identity implied by additive effects hold while
one input is varied and the other three are held fixed. This forces $h$
to be constant on each cell support; overlap equates the constants, and
$D'\mathbf1_n=0$ makes their common value zero. Continuity on actual
supports suffices, so an additional positive-density assumption is not
needed. Connected marginal supports alone do not establish
Assumption~1; Appendix~\ref{app:primitive_counterexample} gives a
counterexample.

\begin{theorem}[Frontier and error-scale identification]
\label{thm:identification}
Suppose Assumption~1 and \eqref{eq:u_dist} hold, and $D$ has rank $q<n$.
For $k=1,2$, let
$\vartheta_k=(f_k,\delta_k,\sigma_{u,k}^2,\sigma_{v,k}^2)$ with
$f_k\in\mathcal F$, $\sigma_{u,k}^2\geq0$, and $\sigma_{v,k}^2>0$.
\begin{enumerate}[1)]
\item If the two configurations generate the same conditional law of
$R'Y$ given $X$, then
\[
f_1=f_2\quad\mu_X\text{-almost everywhere},\qquad
\sigma_{u,1}^2=\sigma_{u,2}^2,\qquad
\sigma_{v,1}^2=\sigma_{v,2}^2.
\]
\item If they generate the same conditional law of the full vector $Y$
given $X$, then also $\delta_1=\delta_2$.
\end{enumerate}
Thus the frontier and the two error-scale parameters are identified by
the invariant distribution. The normalized fixed effects are identified
by the full distribution, although they are absent from the invariant
likelihood.
\end{theorem}

\noindent{\em Proof}. See Appendix~\ref{app:proof_identification}. 

The variance argument remains valid even when a contrast has zero
skewness. For any nonzero $a\in\mathcal C(D)^\perp$,
\begin{align}
\kappa_2(a'Y\mid X)
&=\left[\sigma_v^2+\left(1-\frac2\pi\right)\sigma_u^2\right]
  \sum_{j=1}^n a_j^2,
\label{eq:contrast_kappa2}\\
\kappa_4(a'Y\mid X)
&=\frac{8(\pi-3)}{\pi^2}(\sigma_u^2)^2
  \sum_{j=1}^n a_j^4.
\label{eq:contrast_kappa4}
\end{align}
The fourth cumulant identifies the nonnegative quantity $\sigma_u^2$;
the second then identifies $\sigma_v^2$. For a double difference
$Y_{it}-Y_{is}-Y_{jt}+Y_{js}$, the sums in
\eqref{eq:contrast_kappa2} and \eqref{eq:contrast_kappa4} both equal four.
These are population identification statements, not claims that sample
fourth cumulants necessarily yield precise estimates. Identification at
$\sigma_u^2=0$ also does not imply regular inference at that boundary.

\begin{corollary}[Identification of frontier derivatives]
\label{cor:derivative_identification}
Under the conditions of Theorem~\ref{thm:identification}, suppose an open set
$\mathcal X_0$ is contained in $\operatorname{supp}(\mu_X)$ and every
$f\in\mathcal F$ is continuously differentiable on $\mathcal X_0$.
Then $f(x)$ and $\nabla f(x)$ are identified at every
$x\in\mathcal X_0$. When $x_k$ is log input $k$, $\partial f/\partial x_k$
is its output elasticity. Summing these elasticities gives returns to scale.
These quantities are therefore identified on $\mathcal X_0$.
\end{corollary}

\noindent{\em Proof}. See Appendix~\ref{app:proof_derivative_identification}. 

If Assumption~1 fails, invariant data identify the class of frontiers
satisfying $M_D(f_1-f_2)_X=0$ almost surely. A particular derivative is
identified only if it is unchanged across that class. A GP prior can
regularize a choice within the class but cannot establish structural
identification by itself.

\subsection{Gaussian process representation}
\label{subsec:gp}

Place a GP prior on the common frontier,
\begin{equation*}
f(\cdot)\mid\theta\sim GP\{0,k_\theta(\cdot,\cdot)\},
\end{equation*}
independently of $u$ and $v$ conditional on $X$ and the parameters. At the
observed inputs,
\begin{equation}
\label{eq:fX_prior}
f_X\mid X,\theta\sim N(0,K_\theta),\qquad
K_\theta=[k_\theta(x_{it},x_{js})]_{(it),(js)}.
\end{equation}
A baseline specification is the automatic relevance determination (ARD)
squared-exponential kernel
\begin{equation}
\label{eq:SE}
k_\theta(x,x')=\sigma_f^2
\exp\left\{-\frac12\sum_{k=1}^p
\frac{(x_k-x_k')^2}{\ell_k^2}\right\},
\quad \sigma_f^2>0,\quad \ell_k>0,
\end{equation}
with $\theta=(\sigma_f^2,\ell_1,\ldots,\ell_p)'$. Input-specific length
scales allow different rates of frontier variation across inputs.

\begin{proposition}[GP projection and completion]
\label{prop:gp_completion}
Let $\eta=R'f_X$ and suppose $K_{\theta,R}=R'K_\theta R$ is positive
definite. Then
\begin{equation}
\label{eq:eta_prior}
\eta\mid X,\theta\sim N(0,K_{\theta,R}),
\end{equation}
and
\begin{equation}
\label{eq:f_given_eta}
f_X\mid\eta,X,\theta\sim N(m_f,V_f),
\end{equation}
where $m_f=K_\theta R K_{\theta,R}^{-1}\eta$ and 
$V_f=K_\theta-K_\theta R K_{\theta,R}^{-1}R'K_\theta$.
In particular, $R'm_f=\eta$, $R'V_f=0$, and $V_f$ is positive
semidefinite. If $K_\theta$ is positive definite, then
$K_{\theta,R}$ is positive definite and $\operatorname{rank}(V_f)=q$.
Under the eliminated likelihood below, conditioning additionally on
$(y,u,\sigma_v^2)$ does not change \eqref{eq:f_given_eta}.
\end{proposition}

\noindent{\em Proof}. See Appendix~\ref{app:proof_gp_completion}. 

According to Proposition~\ref{prop:gp_completion}, $\eta$ contains the realized sample information about $f_X$ used by
the invariant likelihood. The structural frontier is still $f$;
it is not redefined as the vector $R\eta$. The completion in
\eqref{eq:f_given_eta} describes GP regularization in the remaining
directions. It does not replace Assumption~1. Positive definiteness of
$K_{\theta,R}$ is needed for these inverse formulas; repeated inputs may
make $K_\theta$ singular even under \eqref{eq:SE}.

\subsection{Fixed-effect-eliminated likelihood}
\label{subsec:concentration}

Let $e=y-f_X+u$. Conditional on $(f_X,u)$, the full Gaussian likelihood is
\begin{equation}
\label{eq:full_ll}
L(y\mid\delta,f_X,u,\sigma_v^2)
=(2\pi\sigma_v^2)^{-n/2}
\exp\{-\|e-D\delta\|^2/(2\sigma_v^2)\}.
\end{equation}
The invariant observation equation is
\begin{equation*}
R'y=\eta-R'u+R'v,\qquad R'v\sim N(0,\sigma_v^2I_r).
\end{equation*}
Its likelihood is the proper density of $w=R'y$,
\begin{equation}
\label{eq:conc_eta}
L_E(y\mid\eta,u,\sigma_v^2)
=(2\pi\sigma_v^2)^{-r/2}
\exp\left\{-\frac{\|R'(y+u)-\eta\|^2}{2\sigma_v^2}\right\}.
\end{equation}
Completing the square in $\delta$ gives
\begin{equation}
\label{eq:conditional_integrated_likelihood}
\int_{\mathbb R^q}L(y\mid\delta,f_X,u,\sigma_v^2)\,d\delta
=|D'D|^{-1/2}L_E(y\mid\eta,u,\sigma_v^2).
\end{equation}
Lebesgue measure here is Haar measure for additive fixed-effect
translations. The integrated expression is a likelihood in the model
parameters; it is not a probability density on all of $\mathbb R^n$.
By contrast, \eqref{eq:conc_eta} is a normalized density on the contrast
space. The determinant factor is parameter free because $D$ is fixed.

The ordinary profile likelihood evaluates \eqref{eq:full_ll} at
$\widehat\delta=(D'D)^{-1}D'e$. It has the same quadratic form but retains
the normalizing factor $(2\pi\sigma_v^2)^{-n/2}$. After integration,
the exponent is $-r/2$, where $r=n-q$ is the residual dimension. This is the error-contrast construction of
restricted likelihood \citep{PattersonThompson1971,Harville1974,Harville1977};
its relation to profile-score adjustment is made explicit below
\citep{CoxReid1987}.

\subsection{Covariance scores after GP integration}
\label{subsec:score_adjustment}

Integrating $f_X$ under \eqref{eq:fX_prior}, conditional on $u$, gives
\begin{equation}
\label{eq:marginal_gp_model}
z(u)=y+u\mid\delta,u,X,\omega\sim N(D\delta,\Sigma_\omega),
\qquad \Sigma_\omega=K_\theta+\sigma_v^2I_n,
\end{equation}
where $\omega=(\sigma_v^2,\sigma_f^2,\rho_1,\ldots,\rho_p)'$ and
$\rho_k=\log\ell_k$. Define
\begin{align*}
A_\omega=D'\Sigma_\omega^{-1}D,\qquad
P_\omega=\Sigma_\omega^{-1}
-\Sigma_\omega^{-1}DA_\omega^{-1}D'\Sigma_\omega^{-1},
\qquad \Omega_\omega=R'\Sigma_\omega R.
\end{align*}
The following results are restricted Gaussian likelihood identities
\citep{Harville1977}. Their application here supplies covariance updates
for the GP frontier conditional on the latent one-sided errors.

\begin{theorem}[Elimination and covariance-score adjustment]
\label{thm:score_adjustment}
Suppose $D$ has rank $q<n$ and is parameter independent. Suppose
$\Sigma_\omega$ is positive definite and twice continuously
differentiable on an open parameter set. Write
$\Sigma_j=\partial\Sigma_\omega/\partial\omega_j$.
\begin{enumerate}[1)]
\item The density of $R'z(u)$ is
\begin{align}
L_E(\omega;y,u)
&=(2\pi)^{-r/2}|\Omega_\omega|^{-1/2}
  \exp\{-z(u)'P_\omega z(u)/2\}
\label{eq:general_eliminated_likelihood}\\
&=(2\pi)^{-r/2}|D'D|^{1/2}|\Sigma_\omega|^{-1/2}
  |A_\omega|^{-1/2}\exp\{-z(u)'P_\omega z(u)/2\}.
\notag
\end{align}
In particular,
$P_\omega=R\Omega_\omega^{-1}R'$ and $|\Sigma_\omega||A_\omega|=|D'D||\Omega_\omega|$.
The integral over $\delta$ of the full Gaussian likelihood is
$|D'D|^{-1/2}L_E$.

\item The eliminated score is
\begin{equation}
\label{eq:eliminated_score}
S_{E,j}=\frac12z(u)'P_\omega\Sigma_jP_\omega z(u)
-\frac12\operatorname{tr}(P_\omega\Sigma_j).
\end{equation}
If $S_{P,j}$ is the ordinary profile score, then
\begin{equation*}
S_{E,j}=S_{P,j}+A_j(\omega),
\end{equation*}
where
\begin{align}
\label{eq:score_adjustment_term}
A_j(\omega)
=\tfrac12\operatorname{tr}\{(\Sigma_\omega^{-1}-P_\omega)\Sigma_j\}
=\tfrac12\operatorname{tr}
\{A_\omega^{-1}D'\Sigma_\omega^{-1}\Sigma_j\Sigma_\omega^{-1}D\}.
\end{align}

\item Under the Gaussian model \eqref{eq:marginal_gp_model}, at its
generating value $\omega_0$,
\begin{align}
E_{\omega_0}(S_{E,j}\mid u,X)&=0,
\label{eq:score_centering}\\
E_{\omega_0}(S_{P,j}\mid u,X)&=-A_j(\omega_0).
\label{eq:profile_score_bias}
\end{align}
The expected information, evaluated at the parameter generating that
Gaussian law, is
\begin{equation}
\label{eq:expected_information}
\mathcal I_{E,jk}(\omega)
=\tfrac12\operatorname{tr}(P_\omega\Sigma_jP_\omega\Sigma_k).
\end{equation}

\item The likelihood, scores, and information are unchanged if $R$ is
replaced by another orthonormal basis of $\mathcal C(D)^\perp$.

\item At an interior point $\omega_0$, let
$\Omega_j=R'\Sigma_jR$ be evaluated at $\omega_0$. The information matrix
is positive definite if and only if these projected derivative matrices
are linearly independent. This rank condition is sufficient for local
identification of $\omega$ from the conditional Gaussian contrast law.
\end{enumerate}
\end{theorem}

\noindent{\em Proof}. See Appendix~\ref{app:proof_score_adjustment}. 

The expectations in this theorem average over both the GP frontier and
Gaussian noise, with $u$ and $X$ fixed. They are not assertions of score
unbiasedness for every fixed structural frontier. Nor does positive
conditional information establish global identification of kernel
parameters or their precision when $u$ is unobserved. The observed-data
score below accounts for that latent uncertainty.

\begin{corollary}[Scores for input-specific length scales]
\label{cor:ard_bandwidth_scores}
For \eqref{eq:SE}, write
\begin{equation*}
[\Delta_k]_{ab}=(x_{a,k}-x_{b,k})^2,\qquad k=1,\ldots,p,
\end{equation*}
where $a,b$ index stacked observations and $\odot$ denotes the Hadamard
product. Then
\begin{equation}
\label{eq:Sigma_sigv_ard}
\Sigma_{\sigma_v^2}=I_n,\qquad
\Sigma_{\sigma_f^2}=K_\theta/\sigma_f^2,\qquad
\Sigma_{\rho_k}=K_\theta\odot(\Delta_k/\ell_k^2).
\end{equation}
The score for $\rho_k$ is
\begin{align}
S_{E,\rho_k}
=\tfrac12z(u)'P_\omega
   \{K_\theta\odot(\Delta_k/\ell_k^2)\}P_\omega z(u)
-\tfrac12\operatorname{tr}
   [P_\omega\{K_\theta\odot(\Delta_k/\ell_k^2)\}].
\label{eq:ard_score}
\end{align}
Under the model in Theorem~\ref{thm:score_adjustment},
\begin{equation*}
E_{\omega_0}\{S_{E,\rho_k}(\omega_0)\mid u,X\}=0,
\qquad k=1,\ldots,p,
\end{equation*}
and
\begin{equation}
\label{eq:ard_information}
\mathcal I_{E,\rho_k\rho_m}
=\tfrac12\operatorname{tr}
 (P_\omega\Sigma_{\rho_k}P_\omega\Sigma_{\rho_m}).
\end{equation}
Joint regular local identification of all covariance parameters follows
if the $p+2$ projected matrices in \eqref{eq:Sigma_sigv_ard} are linearly
independent. In particular, length-scale derivatives must be independent
modulo the span of the projected noise and signal derivatives, and the
two scale derivatives must themselves be independent. If
$\lambda_k=\ell_k^{-2}$ is used instead, then
$\partial K_\theta/\partial\lambda_k=-\tfrac12K_\theta\odot\Delta_k$.
\end{corollary}

\noindent{\em Proof}. See Appendix~\ref{app:proof_ard_corollary}. 

\begin{corollary}[Conditional noise-variance score]
\label{cor:conditional_scores}
Condition on $(u,\eta,X)$ and set
$H=\|R'(y+u)-\eta\|^2$. Then
\begin{equation}
\label{eq:sigma_v_score_E}
S_E(\sigma_v^2)=-\frac{r}{2\sigma_v^2} +\frac{H}{2(\sigma_v^2)^2},~~\text{ and }~~
S_E(\sigma_v^2)=S_P(\sigma_v^2)+\frac{q}{2\sigma_v^2}.
\end{equation}
At the generating value, $H/\sigma_v^2\sim\chi_r^2$. Consequently the
conditional expected eliminated score is zero, and the expected profile
score is $-q/(2\sigma_v^2)$. With an independent inverse-gamma prior
$p(\sigma_v^2)\propto(\sigma_v^2)^{-a_v-1}\exp(-b_v/\sigma_v^2)$,
\begin{equation}
\label{eq:sigma_v_full_conditional}
\sigma_v^2\mid y,u,\eta,X,\theta
\sim IG(a_v+r/2,\ b_v+H/2).
\end{equation}
\end{corollary}

\noindent{\em Proof}. See Appendix~\ref{app:proof_conditional_scores}.

The corresponding oracle estimators, with $(u,\eta)$ supplied at their
generating values, are $H/r$ for the eliminated likelihood and $H/n$ for
the profile likelihood. For fixed $T$ and $N\to\infty$, they converge to
$\sigma_v^2$ and $\{(T-1)/T\}\sigma_v^2$, respectively. This comparison
concerns the conditional Gaussian problem; it does not establish an
asymptotic limit for an estimator that also estimates the unknown
frontier and inefficiencies.

\subsection{Latent inefficiency and observed-data scores}
\label{subsec:ineff_projection}

The vector $R'u$ can have negative elements and dependent coordinates.
It is not an inefficiency vector. Retain the original observation-level
inefficiencies, with
\begin{equation}
\label{eq:u_joint}
p(u\mid s_u)=\left(\frac{2}{\pi s_u}\right)^{n/2}
\exp\left(-\frac{u'u}{2s_u}\right)\mathbf1\{u\in\mathbb R_+^n\},
\qquad s_u=\sigma_u^2>0,
\end{equation}
and define technical efficiency as $TE_{it}=\exp(-u_{it})$. Identification
of a common inefficiency distribution does not make each realized
$u_{it}$ observable; individual efficiency statements remain posterior
inferences under the maintained model.

Let $\psi=(\omega',s_u)'$, let $w=R'y$, and use the normalized contrast
likelihood in \eqref{eq:general_eliminated_likelihood}. The observed-data
density on contrast space is
\begin{equation*}
L_O(\psi;w,X)=\int_{\mathbb R_+^n}
L_E(\omega;y,u)\,p(u\mid s_u)\,du.
\end{equation*}
Although written using $y$, the integrand depends on it only through
$w$. Define the complete-data score
\begin{equation}
\label{eq:complete_score}
S_C(\psi;w,u)=
\begin{pmatrix}
S_E(\omega;y,u)\\
-n/(2s_u)+u'u/(2s_u^2)
\end{pmatrix}.
\end{equation}

\begin{proposition}[Observed-data score and information]
\label{prop:observed_score}
Suppose $s_u>0$, differentiation under the relevant integrals is valid
through second order, and the scores have finite second moments. Under
the GP-integrated model,
\begin{equation}
\label{eq:fisher_identity}
S_O(\psi;w,X)=E_\psi\{S_C(\psi;w,U)\mid W=w,X\}.
\end{equation}
Here the conditional law of $U$ has density proportional to
$L_E(\omega;y,u)p(u\mid s_u)$ at fixed $\psi$. At the generating parameter,
$E_{\psi_0}\{S_O(\psi_0;W,X)\mid X\}=0$. Moreover, with
$J_C=-\nabla_\psi^2\log\{L_Ep(U\mid s_u)\}$,
\begin{align}
J_O(\psi;w,X)
&=E_\psi(J_C\mid w,X)
  -\operatorname{Var}_\psi(S_C\mid w,X),
\label{eq:louis_identity}\\
\mathcal I_O(\psi\mid X)
&=\mathcal I_C(\psi\mid X)
 -E_\psi\{\operatorname{Var}_\psi(S_C\mid W,X)\mid X\}
 \preceq\mathcal I_C(\psi\mid X),
\label{eq:missing_information}
\end{align}
where $J_O=-\nabla_\psi^2\log L_O$,
$\mathcal I_O=E_\psi(J_O\mid X)$, and
$\mathcal I_C=\operatorname{Var}_\psi(S_C\mid X)$.
\end{proposition}

\noindent{\em Proof}. See Appendix~\ref{app:proof_observed_score}. 

The score identity and the observed-information decomposition follow
from latent-variable likelihood differentiation \citep{Louis1982}.
They permit calculation using conditional draws of $u$ at fixed
parameters. The missing-information term explains why the Gaussian
information conditional on $u$ is not the information available when
inefficiency is unobserved. At $s_u=0$, \eqref{eq:u_joint} is replaced by
a point mass; the interior score formulas for $s_u$ do not apply.

\section{Bayesian estimation and posterior inference}
\label{sec:bayes}

Posterior inference uses the fixed-effect-eliminated likelihood in
\eqref{eq:conc_eta}. The sampling state contains the projected frontier
$\eta=R'f_X$, the original observation-level inefficiencies, and the
variance and kernel parameters. Frontier values, fixed effects, and
derivatives are recovered from retained draws. Throughout this section,
conditioning on $X$ is implicit, $K_R=K_{\theta,R}$ is assumed positive
definite, and
\[
s_f=\sigma_f^2,\qquad s_v=\sigma_v^2,\qquad s_u=\sigma_u^2,
\qquad z_u=R'(y+u),\qquad \Omega_\theta=K_R+s_vI_r.
\]
The notation $z_u$ emphasizes that this quantity changes whenever $u$
is updated. We generally use Cholesky factorizations and triangular
solves for covariance systems. In the dairy application, we use a
relative eigenvalue cutoff only when recovering derivatives from retained
posterior draws. For the eigenvalues $d_j$ of $(K_R+K_R')/2$, the solve
assigns weight $1/d_j$ to directions with $d_j>10^{-10}\max_k d_k$
and zero weight to the rest.

\subsection{Priors and posterior}
\label{subsec:priors}

Retain the Gaussian prior for $\eta$ in \eqref{eq:eta_prior}. For the
ARD kernel, assign independent proper priors
\begin{equation}
s_f\sim G\left(\frac12,\frac{1}{2\xi_f}\right),\qquad
\ell_k^{-2}\sim
G\left(\frac12,\frac{1}{2\xi_{\ell,k}}\right),
\qquad k=1,\ldots,p,
\label{eq:post_prior_lengthscales}
\end{equation}
where $G(a,b)$ has shape $a$ and rate $b$, and all $\xi$ parameters
are strictly positive. The use of scale-adapted GP priors follows the
general motivation of \citet{hauzenberger2025}; the input-specific
specification above is stated explicitly for the present model.

Prior scales are fixed before sampling. Input-distance calibration and
the basis-invariant response calibration are described in
Appendix~\ref{app:prior_calibration}; using the response gives an
empirical Bayes analysis conditional on the selected scale.

For the two squared error scales, take independent priors
\[
s_v\sim IG(a_v,b_v), ~~\text{ and }~~
s_u\sim IG(a_u,b_u),
\]
with $a_v,b_v,a_u,b_u>0$. The convention is
$p(s)=b^a\{\Gamma(a)\}^{-1}s^{-a-1}\exp(-b/s)$ for $s>0$.
These continuous priors assign probability one to positive scales;
they do not assign posterior mass to the boundary $s_u=0$.

For $\Theta=(\eta,u,s_v,s_u,\theta)$, the posterior is
\begin{equation}
\label{eq:post_joint}
\pi_E(\Theta\mid y)
\propto L_E(y\mid\eta,u,s_v)\,p(\eta\mid\theta)
\,p(u\mid s_u)\,p(s_v)\,p(s_u)\,p(\theta).
\end{equation}
Here $p(u\mid s_u)$ is the half-normal product density in
\eqref{eq:u_joint}, and $p(\theta)$ is the density induced by
\eqref{eq:post_prior_lengthscales}
under $\theta=(s_f,\ell_1,\ldots,\ell_p)'$. No fixed-effect
coefficient appears in this posterior. All original $u_{it}$ remain
in the state, so observation-level efficiency is retained.

The proper priors above yield a proper posterior for every fixed choice
of their positive scales; Appendix~\ref{app:posterior_propriety} gives
the integrability argument.

\subsection{Conditional posterior updates}
\label{subsec:eta_post}

Given $u$, the Gaussian observation equation is
$z_u=\eta+\varepsilon_R$, where
$\varepsilon_R\sim N(0,s_vI_r)$. Completing the square with the
GP prior gives
\begin{equation}
\label{eq:post_eta_cond}
\eta\mid u,\theta,s_v,y\sim N(\bar\eta,V_\eta),
\end{equation}
where the equivalent covariance and mean expressions are
\begin{align*}
V_\eta&=(K_R^{-1}+s_v^{-1}I_r)^{-1}
       =K_R-K_R\Omega_\theta^{-1}K_R,\\
\bar\eta&=s_v^{-1}V_\eta z_u
          =K_R\Omega_\theta^{-1}z_u.
\end{align*}
An exact Gaussian simulation that avoids subtracting covariance matrices
is given in Appendix~\ref{app:conditional_simulation}.

For the inefficiency block, define $r_\eta=y-R\eta$. The conditional
kernel is
\begin{equation*}
\pi_E(u\mid\eta,s_u,s_v,y)
\propto
\exp\left\{-\frac{(r_\eta+u)'M_D(r_\eta+u)}{2s_v}
            -\frac{u'u}{2s_u}\right\}
\mathbf1\{u\geq0\}.
\end{equation*}
Consequently,
\begin{equation}
\label{eq:post_u_tmvn}
u\mid\eta,s_u,s_v,y\sim TN_n^+(m_u,V_u),
\end{equation}
where $TN_n^+$ denotes a Gaussian law restricted to the positive
orthant and
\begin{align*}
V_u^{-1}=s_u^{-1}I_n+s_v^{-1}M_D, \quad
V_u=s_uI_n-\frac{s_u^2}{s_u+s_v}M_D,\quad
m_u=-\frac{s_u}{s_u+s_v}M_Dr_\eta.
\end{align*}
The inverse follows from $M_D^2=M_D$; the mean follows by
completing the square. These are the parameters of the untruncated
Gaussian kernel, rather than the moments after truncation.

The inefficiencies can be updated by a joint truncated-normal draw
or by complete coordinate Gibbs sweeps. The coordinate distributions
and an implementation using row and column sums are given in
Appendix~\ref{app:conditional_simulation}. A sweep preserves
\eqref{eq:post_u_tmvn}, but need not produce an independent joint draw.

Let
$SSR_E=\|z_u-\eta\|^2=(r_\eta+u)'M_D(r_\eta+u)$.
Multiplication by the inverse-gamma priors yields
\begin{align}
s_v\mid\eta,u,\theta,s_u,y
&\sim IG\left(a_v+\frac r2,\ b_v+\frac12SSR_E\right),
\label{eq:post_sigv}\\
s_u\mid u,\eta,\theta,s_v,y
&\sim IG\left(a_u+\frac n2,\ b_u+\frac12u'u\right).
\label{eq:post_sigu}
\end{align}
The first shape parameter uses the $r$ Gaussian contrast coordinates;
the second uses the $n$ original half-normal inefficiencies. Eliminating
fixed effects does not replace $n$ by $r$ in the inefficiency update.

\subsection{Partially collapsed posterior simulation}
\label{subsec:algorithm}

Integrating out $\eta$ gives the collapsed conditional target
\[
\pi_E(\theta\mid u,s_v,y)\propto
|\Omega_\theta|^{-1/2}
\exp\{-z_u'\Omega_\theta^{-1}z_u/2\}\,p(\theta).
\]
The random-walk update uses $\phi=(\log s_f,\log\ell_1,\ldots,\log\ell_p)'$.
Appendix~\ref{subsec:gp_hyper} gives its log target, including the
transformation Jacobians, and Appendix~\ref{subsec:ridge_move} describes
an optional joint move for inefficiency and the error scales.

Algorithm~\ref{alg:pcg} combines this collapsed hyperparameter update
with conditional Gaussian, truncated Gaussian, and inverse-gamma
updates. Its ordering follows the partially collapsed Gibbs
construction of \citet{vandyk2008}; Metropolis steps inside such
schemes require preservation of the intended marginal target
\citep{vandykjiao2015}.

\begin{algorithm}[htbp!]
\caption{Partially collapsed Gibbs sampler for the two-way GP frontier}
\label{alg:pcg}
\textit{Setup.} Construct $D$, $R$, and the input distance matrices.
Fix the prior scales and choose initial values
$\phi\in\mathbb R^{p+1}$, $u\in\mathbb R_+^n$, $s_u>0$, and $s_v>0$.
At every step, use the latest accepted state.
\begin{enumerate}[1)]
\item Recompute $z_u=R'(y+u)$. Update $\phi$ by a symmetric
random-walk Metropolis step using the log target
\eqref{eq:post_phi_target}, which integrates out $\eta$. Set
$\theta=\theta(\phi)$.
\item Immediately draw $\eta$ from \eqref{eq:post_eta_cond}, using
the current $(\theta,u,s_v)$. Perform this draw whether the
hyperparameter proposal was accepted or rejected.
\item Update all coordinates of $u$ sequentially using
\eqref{eq:post_uj_cond}, for a fixed positive number of complete
sweeps. An exact joint draw from \eqref{eq:post_u_tmvn} is an
alternative. Recompute $z_u$ after the block.
\item If used, attempt the optional move
\eqref{eq:post_ridge_proposal} with acceptance probability
\eqref{eq:post_ridge_acceptance}. Refresh $z_u$ and the residual
norm after every accepted move.
\item Draw $s_v$ from \eqref{eq:post_sigv} and $s_u$ from
\eqref{eq:post_sigu}, using the current $(u,\eta)$.
\end{enumerate}
\textit{Retained draws.} After warm-up, store the state and obtain
the frontier, fixed effects, efficiency, and derivative summaries
described below. Freeze proposal adaptation before retaining draws.
\end{algorithm}

The invariant target is \eqref{eq:post_joint}. Fix the remaining state $b=(u,s_u,s_v)$ and
factor the conditional posterior as
$\pi_E(\phi,\eta\mid b,y)=\pi_E(\phi\mid b,y)\pi_E(\eta\mid\phi,b,y)$.
Step 1 leaves the first factor invariant and does not use the old
$\eta$; step 2 restores the second factor by an exact draw.
Their composition therefore preserves the joint conditional law.
The subsequent Gibbs sweeps, optional Metropolis move, and variance
updates each preserve their corresponding conditional target.
This argument establishes posterior invariance. The convergence rate
requires a separate analysis.


\subsection{Recovery of the structural frontier and fixed effects}
\label{subsec:frontier_recovery}

The invariant likelihood depends on $f_X$ only through $\eta$.
Proposition~\ref{prop:gp_completion} therefore supplies the posterior
completion in \eqref{eq:f_given_eta} for every retained
$(\eta,\theta)$ draw. To simulate from this potentially singular
Gaussian law, draw $\widetilde f_X\sim N(0,K_\theta)$ and set
\begin{equation*}
f_X=\widetilde f_X+K_\theta R K_R^{-1}
                       (\eta-R'\widetilde f_X).
\end{equation*}
The correction gives $R'f_X=\eta$ exactly. Its mean and covariance
are $m_f$ and $V_f$ in \eqref{eq:f_given_eta}, so uncertainty in the
remaining frontier directions is retained. Treating $R\eta$ as the
complete frontier would discard that uncertainty.

At a new input $x_\star$, let
$k_\star=[k_\theta(x_j,x_\star)]_{j=1}^n$. Gaussian conditioning
also gives
\[
f(x_\star)\mid\eta,\theta
\sim N\left(k_\star'RK_R^{-1}\eta,\,
k_\theta(x_\star,x_\star)-k_\star'RK_R^{-1}R'k_\star\right).
\]
Evaluation at several inputs uses their joint kernel covariance.
Posterior uncertainty combines these conditional Gaussian laws over
the retained parameter draws.

To recover normalized fixed effects, return to the full model under
the same flat measure on $\delta$. Completing the Gaussian square
gives
\begin{equation}
\label{eq:post_delta_distribution}
\delta\mid f_X,u,s_v,y
\sim N\{\widehat\delta,\,s_v(D'D)^{-1}\},
\qquad \widehat\delta=(D'D)^{-1}D'(y-f_X+u).
\end{equation}
Then $\alpha=C_Na$ and $\gamma=C_Tc$ for $\delta=(a',c')'$.
For posterior means, averaging $\widehat\delta$ over the retained
draws integrates out the conditional Gaussian noise. Credible
intervals must also include the covariance $s_v(D'D)^{-1}$, either
analytically or through conditional draws from
\eqref{eq:post_delta_distribution}. These are posterior reconstructions
under the original model and flat nuisance measure; the contrast
likelihood itself contains no fixed-effect coefficient.

\subsection{Efficiency, elasticities, and returns to scale}
\label{subsec:derivatives}

For each retained draw, set $TE_{it}^{(s)}=\exp(-u_{it}^{(s)})$.
Averaging these draws gives the estimated posterior mean efficiency.
Their empirical quantiles give the endpoints of equal-tailed credible intervals.
The squared-exponential kernel also gives conditional Gaussian
laws for frontier derivatives. Define $g(x_\star)=\nabla f(x_\star)$
and the $n\times p$ matrix $B_\star=[d_{1,\star},\ldots,d_{p,\star}]$,
where
\begin{equation*}
[d_{k,\star}]_j
=\frac{\partial k_\theta(x_j,x_\star)}{\partial x_{\star,k}}
=k_\theta(x_j,x_\star)\frac{x_{j,k}-x_{\star,k}}{\ell_k^2}.
\end{equation*}
The derivative covariance follows from
\begin{equation*}
\frac{\partial^2 k_\theta(x,x')}{\partial x_k\partial x'_m}
=k_\theta(x,x')\left\{
\frac{\mathbf1\{k=m\}}{\ell_k^2}
-\frac{(x_k-x'_k)(x_m-x'_m)}{\ell_k^2\ell_m^2}\right\}.
\end{equation*}
At $x=x'=x_\star$, this gives
$C^{(1,1)}_{\star\star}
=\operatorname{diag}(s_f/\ell_1^2,\ldots,s_f/\ell_p^2)$.
Since $\operatorname{Cov}(g(x_\star),\eta)=B_\star'R$,
\begin{equation}
\label{eq:post_gradient_distribution}
g(x_\star)\mid\eta,\theta
\sim N\left(B_\star'RK_R^{-1}\eta,\,
C^{(1,1)}_{\star\star}-B_\star'RK_R^{-1}R'B_\star\right).
\end{equation}
Drawing from \eqref{eq:post_gradient_distribution} at each retained
$(\eta,\theta)$ draw incorporates both conditional frontier
uncertainty and parameter uncertainty. The resulting marginal
posterior is a mixture of Gaussian laws and need not itself be
Gaussian.

For a vector $h$ collecting frontier values and derivatives at several inputs, form its prior
covariance $C_h$ and cross-covariance $C_{h\eta}$ using the kernel
and its derivatives. The appropriate conditional law is
\[
h\mid\eta,\theta\sim
N(C_{h\eta}K_R^{-1}\eta,\,
C_h-C_{h\eta}K_R^{-1}C_{h\eta}').
\]
Independent draws from separate marginal conditional laws would
omit their conditional covariance. Pointwise credible intervals
also do not constitute a simultaneous band over the input space.

If $x_k$ is the unstandardized log of physical input $q_k$, its
elasticity is $g_k(x)$. If $x_k=q_k$ enters in levels, the elasticity
is $q_kg_k(x)$. More generally, for a separately transformed input
$x_k=T_k(q_k)$, apply the chain rule,
\[
\varepsilon_k(q)=q_kT_k'(q_k)\,g_k(T(q)).
\]
For example, if $x_k=(\log q_k-a_k)/b_k$, the elasticity is
$g_k(x)/b_k$. Returns to scale (RTS) are the sum over production inputs,
$\operatorname{RTS}(q)=\sum_k\varepsilon_k(q)$, holding other
controls fixed. Their posterior intervals are computed from joint
elasticity draws. Structural interpretation on the input support requires the
identification conditions in Section~\ref{subsec:identification};
extrapolation and directions left unidentified by those conditions
remain dependent on GP regularization.

\section{Monte Carlo simulation}
\label{sec:simulation}

This section evaluates the finite-sample performance of the proposed
fixed-effect-eliminated Gaussian process stochastic frontier estimator.
For each data-generating process (DGP), the sample is a balanced panel of
$N$ production units observed over $T$ periods. For each configuration,
the individual effects are generated from Gaussian draws with standard
deviation $0.2$; where present, the time effects are generated from
Gaussian draws with standard deviation $0.25$. The draws are then centered
to satisfy the zero-sum normalizations. The resulting effects are held
fixed across Monte Carlo replications, so the reported performance
measures are conditional on these realized effects.

Let $\tilde\alpha_i=\alpha_i/0.2$ and
$\tilde\gamma_t=\gamma_t/0.25$. The two input variables are generated as
\begin{align*}
x_{1,it}
&=0.2\tilde\alpha_i+\omega_1\tilde\gamma_t
  +\sqrt{1-0.2^2-\omega_1^2}\,e_{1,it},\\
x_{2,it}
&=0.8\tilde\alpha_i+\omega_2\tilde\gamma_t
  +\sqrt{1-0.8^2-\omega_2^2}\,e_{2,it},
\end{align*}
where the input innovations $e_{1,it}$ and $e_{2,it}$ are independent
standard normal variables. For the one-way designs,
$(\omega_1,\omega_2)=(0,0)$ and the time-effect terms are absent; for the
two-way designs, $(\omega_1,\omega_2)=(0.6,0.4)$. Thus, the inputs depend
on the individual effects and, in the two-way designs, on the time effects.

Writing $x_{it}=(x_{1,it},x_{2,it})'$, we consider two frontier functions.
The first is a translog frontier,
\[
f_{\mathrm{TL}}(x_{it})
=0.5x_{1,it}+0.4x_{2,it}-0.05x_{1,it}^{2}
 -0.04x_{2,it}^{2}+0.05x_{1,it}x_{2,it},
\]
which is contained in the parametric benchmark. The second is a smooth
nonlinear frontier with an S-shaped component in $x_1$, motivated by the
nonlinear designs of \citet{MartinsFilho2015},
\[
f_{\mathrm{NL}}(x_{it})
=0.5\arctan(2x_{1,it})+0.4x_{2,it}-0.05x_{2,it}^{2}.
\]
This function cannot be represented by a second-order translog
specification. The Gaussian input innovations have unbounded support,
and neither frontier is globally increasing in both inputs under this
design. The simulations therefore evaluate frontier approximation and
fixed-effect handling without imposing global monotonicity.

Combining the two frontier functions with one-way and two-way
heterogeneity, we specify:
\begin{align*}
\text{DGP1:}\quad y_{it}
&=\alpha_i+f_{\mathrm{TL}}(x_{it})-u_{it}+v_{it},\\
\text{DGP2:}\quad y_{it}
&=\alpha_i+f_{\mathrm{NL}}(x_{it})-u_{it}+v_{it},\\
\text{DGP3:}\quad y_{it}
&=\alpha_i+\gamma_t+f_{\mathrm{TL}}(x_{it})-u_{it}+v_{it},\\
\text{DGP4:}\quad y_{it}
&=\alpha_i+\gamma_t+f_{\mathrm{NL}}(x_{it})-u_{it}+v_{it}.
\end{align*}
In every DGP, $u_{it}\sim N^+(0,0.6^2)$ and
$v_{it}\sim N(0,0.1^2)$. The disturbance draws are independent across
observations, mutually independent, and independent of the input
innovations. The true error-scale parameters are therefore
$\sigma_u=0.6$ and $\sigma_v=0.1$.

We fit the four models in Table~\ref{tab:simu_models} to each
simulated dataset. The proposed estimator is denoted by
TWFE--GP. Comparing FE--GP with TWFE--GP assesses the contribution of
time effects while retaining a GP frontier. Comparing TWFE--TL with
TWFE--GP assesses frontier flexibility while retaining both sets of
fixed effects. Each estimator uses the orthogonal-contrast likelihood
appropriate to its maintained fixed effects. For the GP estimators,
posterior simulation follows Section~\ref{sec:bayes}.

\begin{table}[htbp]
\centering
\caption{Competing models used in the simulation study.}
\label{tab:simu_models}
\begin{tabular}{lcc}
\toprule
 & Translog frontier & Gaussian process frontier \\
\midrule
Individual fixed effects
 & FE--TL & FE--GP \\
Individual and time fixed effects
 & TWFE--TL & TWFE--GP \\
\bottomrule
\end{tabular}
\end{table}

For the scale parameters $\sigma_u$ and $\sigma_v$, we report bias
(BIAS) and root mean squared error (RMSE). For frontier values and
technical efficiency, we report averaged bias (ABIAS) and root averaged
mean squared error (RAMSE). We also report the Spearman rank correlation
between estimated and true technical efficiency, averaged over
replications, to assess the ranking of producers.
Tables~\ref{tab:mc_dgp1}--\ref{tab:mc_dgp4} present the results for
DGP1--DGP4, based on $M=500$ replications for each combination of
$N\in\{25,50\}$ and $T\in\{10,20\}$. The comparisons below describe
the reported performance measures; small differences in their values
should not be interpreted as statistically significant without Monte
Carlo uncertainty estimates.

Table~\ref{tab:mc_dgp1} reports the results for DGP1, with a translog
frontier and individual effects only. FE--TL is correctly specified
and has the lowest RMSE for both scale parameters, the lowest RAMSE for
the frontier and technical efficiency, and the highest rank correlation
among the four models in every $(N,T)$ configuration. The biases are
broadly similar across models. Allowing a GP frontier entails a modest
increase in estimation error: the frontier RAMSE of FE--GP exceeds that
of FE--TL by $0.0066$--$0.0091$, while the differences in technical-
efficiency RAMSE and rank correlation are at most $0.0021$ and $0.0053$,
respectively. Performance generally improves as $N$ or $T$ increases, and the
additional flexibility has a small cost for efficiency estimation.

Table~\ref{tab:mc_dgp2} presents the results for DGP2, with a nonlinear
frontier and individual effects only. FE--GP accommodates the nonlinear
frontier and the one-way heterogeneity, and has the lowest RMSE for both
scale parameters, the lowest RAMSE for the frontier and technical
efficiency, and the highest rank correlation in every configuration.
The translog models impose an incorrect functional form: their frontier
RAMSE ranges from $0.1538$ to $0.1688$, whereas that of FE--GP falls
from $0.0939$ at $(25,10)$ to $0.0491$ at $(50,20)$. This approximation
error is accompanied by upward bias in the estimated noise scale.
The translog models overestimate $\sigma_v$ by $0.0696$--$0.0846$,
relative to its true value of $0.1$, while the bias under FE--GP
falls from $0.0288$ to $0.0053$ between these two configurations.
The GP models also have lower technical-efficiency RAMSE and higher
rank correlations. TWFE--GP
performs nearly as well as FE--GP: its frontier RAMSE is at most $0.0051$ higher
and its rank correlation at most $0.0188$ lower.

Table~\ref{tab:mc_dgp3} reports the results for DGP3, with a translog
frontier and both individual and time effects. TWFE--TL is correctly
specified and generally performs best. It has the lowest frontier and
technical-efficiency RAMSE and the highest rank correlation in all four
configurations. The one-way models omit the time effects and have
substantially larger errors under both frontier specifications. Their
frontier RAMSE ranges from $0.1738$ to $0.3376$, their upward bias in
$\sigma_v$ from $0.0806$ to $0.1029$, and their rank correlations from
$0.8010$ to $0.8327$. These errors remain substantial and do not decline
uniformly across sample configurations. Comparisons across configurations
also reflect changes in the realized fixed effects, which are generated
separately for each configuration. TWFE--GP performs nearly as well as TWFE--TL:
its frontier RAMSE is $0.0064$--$0.0147$ higher, while the differences
in technical-efficiency RAMSE and rank correlation are at most $0.0026$
and $0.0064$, respectively. As in DGP1, the additional frontier
flexibility has a modest cost when the translog specification is correct,
whereas omitting time effects leads to substantial errors.

Table~\ref{tab:mc_dgp4} presents the results for DGP4, with a nonlinear
frontier and both individual and time effects. TWFE--GP combines a
flexible frontier with both sets of fixed effects and has the lowest
RMSE for both scale parameters, the lowest RAMSE for the frontier and
technical efficiency, and the highest rank correlation in all four
configurations. TWFE--TL includes time effects but imposes a translog
frontier; its frontier RAMSE remains between $0.1443$ and $0.1706$.
FE--TL combines frontier misspecification with omitted time effects,
while FE--GP omits time effects despite allowing a flexible frontier.
The frontier RAMSE of these one-way models ranges from $0.1761$ to
$0.3806$. All three restricted models also overestimate the noise scale.
Under TWFE--GP, the bias of $\sigma_v$ falls from $0.0332$ at
$(25,10)$ to $0.0065$ at $(50,20)$. Technical-efficiency RAMSE ranges from $0.0775$
to $0.1000$ under TWFE--GP, compared with $0.1072$--$0.1393$ for the
other models. Its rank correlation ranges from $0.8652$ to $0.9159$,
compared with a maximum of $0.8353$ for TWFE--TL and $0.8237$ for
FE--GP.

The true fixed-effects model of \citet{GREENE2005} estimates the
individual effects jointly with the frontier coefficients and error
scales by maximum likelihood. Joint estimation of an increasing number
of individual effects can create an incidental-parameters problem when
$T$ is fixed \citep{CHEN201465,LANCASTER2000}. We apply this
dummy-variable maximum likelihood estimator, denoted by TFE--MLE, to
DGP1 and compare it with FE--TL in Table~\ref{tab:tfe}. Both use the
correct translog specification, but FE--TL uses the fixed-effect-
eliminated likelihood. TFE--MLE severely underestimates $\sigma_v$:
its bias ranges from $-0.1000$ to $-0.0994$ for $T=10$ and from
$-0.0802$ to $-0.0783$ for $T=20$, against a true value of $0.1$.
The downward bias persists when $N$ increases from $25$ to $50$,
a pattern consistent with incidental-parameter distortion. The Gaussian
factor $(T-1)/T$ following Corollary~\ref{cor:conditional_scores} applies to an oracle variance
estimator; it does not provide a bias formula for the feasible estimator
of $\sigma_v$ in this jointly estimated frontier model. FE--TL has a
smaller absolute bias in $\sigma_v$, which decreases as either $N$ or
$T$ increases over the reported configurations. The comparison is not
uniform across parameters: TFE--MLE has a lower RMSE for $\sigma_u$
when $T=10$, although FE--TL has lower frontier and technical-efficiency
RAMSE and higher rank correlations in all four configurations.

Across these designs, TWFE--GP performs nearly as well as the best-performing
model when a translog frontier or the absence of time effects is correctly
imposed, and performs best when the frontier is nonlinear and time effects
are present. For each value of $T$, doubling $N$ reduces its reported
bias in $\sigma_v$ by approximately half or more in every design.
These finite-sample improvements do not establish fixed-$T$ consistency.
Corollary~\ref{cor:conditional_scores} concerns an oracle conditional
Gaussian calculation and does not establish consistency of the feasible
estimator with an unknown frontier and latent inefficiencies. Within the
designs considered, the results support TWFE--GP when both frontier shape
and time heterogeneity are uncertain.

\section{Technical efficiency and returns to scale on dairy farms in \\ Asturias}
\label{sec:application}

\subsection{Dairy-farm data and competing frontier models}
\label{subsec:application_data}

We apply the model to dairy farms in Asturias, northern Spain, observed
annually from 1999 to 2010. The application examines how frontier
flexibility and the treatment of time heterogeneity affect estimates of
technical efficiency, input elasticities, and returns to scale. These
data extend the sample studied by \citet{ALVAREZ2004241}, who analyzed
196 farms over six years, and have been used extensively in stochastic
frontier research
\citep{amsler2016,Amsler2019,PEREZMENDEZ2020,parmeter2025}.

The extended dataset contains 137 farms. Following \citet{Amsler2019},
we retain farms with complete, strictly positive observations on all
variables required by our specification in every year. The resulting
balanced panel contains 90 farms observed over 12 years, giving 1,080
farm-year observations. This sample permits the explicit fixed-effect
projection for the full $N\times T$ design in Section~\ref{sec:model}.
The empirical findings therefore concern the selected balanced sample.

Output is annual milk production in liters (Milk). The five inputs are
labor measured in full-time equivalents (Labor), the number of milking
cows (Cows), purchased feed for the dairy herd in kilograms (Feed),
agricultural land in hectares (Land), and expenditure associated with
on-farm forage production (Roughage). Roughage is measured as the sum
of expenditure on fertilizer and seeds and the depreciation of machinery
and buildings. It is thus an expenditure proxy for inputs used to
produce forage on the farm. Output and all five inputs enter the model
in logarithms.

Table~\ref{table:descriptivestats} reports descriptive statistics.
Annual milk output ranges from 43,090 to 1,322,276 liters, a ratio of
approximately 30.7, and herd size ranges from 11 to 134 cows. Farm effects
allow for persistent additive differences between farms, while year
effects accommodate additive shifts common to all farms in a given year.
The GP frontier allows input elasticities to vary across the input space
without imposing the constant elasticities of a Cobb--Douglas frontier
or the affine dependence on log inputs implied by a translog frontier.

The proposed specification is
\begin{equation}
\label{eq:sdf}
\log Y_{it}=\alpha_i+\gamma_t+f(\log\bm{x}_{it})-u_{it}+v_{it},
\qquad i=1,\ldots,N,\quad t=1,\ldots,T,
\end{equation}
where $Y_{it}$ is milk output and
$\log\bm{x}_{it}=(\log x_{1,it},\ldots,\log x_{5,it})'$ collects
the log inputs in the order labor, cows, feed, land, and roughage.
The unknown frontier $f$ has the GP prior described in
Section~\ref{sec:model}. The farm and year effects are unrestricted
apart from $\sum_i\alpha_i=\sum_t\gamma_t=0$, so that $f$ contains
the common frontier level. Conditional on the inputs, the noise and
inefficiency components satisfy
\[
v_{it}\stackrel{\mathrm{iid}}{\sim}N(0,\sigma_v^2),
\qquad
u_{it}\stackrel{\mathrm{iid}}{\sim}N^+(0,\sigma_u^2),
\]
with the two arrays mutually independent and their conditional
distributions independent of the input values. No independence
restriction is imposed between the fixed effects and the inputs.
This allowance does not relax the maintained exogeneity of transient
inefficiency and noise.

We compare model~\eqref{eq:sdf}, denoted by TWFE--GP, with two
translog specifications. The two-way fixed-effects benchmark,
denoted by TWFE--TL, is
\[
\log Y_{it}
=\alpha_i+\gamma_t+\beta_0
   +\sum_{j=1}^{5}\beta_j\log x_{j,it}+\frac12\sum_{j=1}^{5}\sum_{k=1}^{5}
      \beta_{jk}\log x_{j,it}\log x_{k,it}-u_{it}+v_{it},
\]
where $\beta_{jk}=\beta_{kj}$. It retains the same additive farm
and year effects as TWFE--GP; the comparison therefore changes the
frontier specification while preserving the fixed-effect structure.

The one-way benchmark, denoted by FE--TL, is
\begin{align*}
\log Y_{it}
={}&\alpha_i+\beta_0+\sum_{j=1}^{5}\beta_j\log x_{j,it}
   +\frac12\sum_{j=1}^{5}\sum_{k=1}^{5}
      \beta_{jk}\log x_{j,it}\log x_{k,it}\\
 &+\beta_t t+\frac12\beta_{tt}t^2
   +\sum_{j=1}^{5}\beta_{jt}t\log x_{j,it}-u_{it}+v_{it},
\end{align*}
again with $\beta_{jk}=\beta_{kj}$ and $\sum_i\alpha_i=0$.
This specification follows the true fixed-effects framework of
\citet{GREENE2005}. Its quadratic trend represents neutral technical
change, and the interactions between time and the log inputs allow
non-neutral technical change. Comparing FE--TL with TWFE--TL therefore
changes both the restrictions on common annual shifts and the allowance
for time-varying input elasticities. The comparison does not isolate
the contribution of unrestricted year effects alone. All three models
use the same normal--half-normal composed-error specification.

\subsection{Error scales and technical efficiency}
\label{subsec:application_efficiency}

For each model, we run the sampler for 110,000 iterations. We discard the
first 10,000 iterations as burn-in and retain every tenth subsequent draw,
leaving 10,000 posterior draws.
Table~\ref{estimate} reports posterior means and 95\% credible intervals
for the main quantities of interest. Posterior summaries of the translog
coefficients are reported in Table~\ref{tab:translog_coef} of
Appendix~\ref{app:translog}.

The estimated inefficiency scales are similar across specifications.
The posterior mean of $\sigma_u$ is $0.0657$ under TWFE--GP,
$0.0651$ under TWFE--TL, and $0.0634$ under FE--TL. The corresponding
posterior means of the noise standard deviation $\sigma_v$ are
$0.0545$, $0.0605$, and $0.0627$. The lower noise estimate under
TWFE--GP is consistent with the flexible frontier accounting for
variation allocated to noise by the translog specifications. It does
not, by itself, establish that the GP frontier is correctly specified.

Table~\ref{estimate} also reports the widely applicable information
criterion (WAIC), which estimates predictive accuracy from posterior
log-likelihood contributions
\citep{Watanabe2010,VehtariGelmanGabry2017}. The reported values are
$-2588.5$ for TWFE--GP, $-2484.4$ for TWFE--TL, and $-2449.7$
for FE--TL. On the conventional deviance scale, lower values indicate
better estimated predictive performance when the scores use a common
predictive target. This qualification is essential here: eliminating
different fixed-effect structures produces different contrast spaces,
so a pointwise WAIC calculated directly from model-specific contrast
coordinates need not be comparable across models. A predictive
comparison requires common observation units and a compatible
treatment of latent quantities. Until these features of the
calculation are verified, the reported values remain provisional
diagnostics rather than evidence establishing a predictive ranking.

Technical efficiency is $TE_{it}=\exp(-u_{it})$. Posterior mean scores
are obtained by averaging $\exp(-u_{it}^{(s)})$ over the retained
draws, as described in Section~\ref{subsec:derivatives}.
The sample means of these scores are $0.9499$, $0.9504$, and $0.9516$
under TWFE--GP, TWFE--TL, and FE--TL, respectively; the corresponding
minimum scores are $0.7839$, $0.7921$, and $0.8071$.
These estimates measure transient technical efficiency under the
maintained specification. Persistent differences absorbed by the
farm effects are not separately identified as inefficiency.

\begin{figure}[h]
    \centering
    \includegraphics[width=0.8\textwidth]{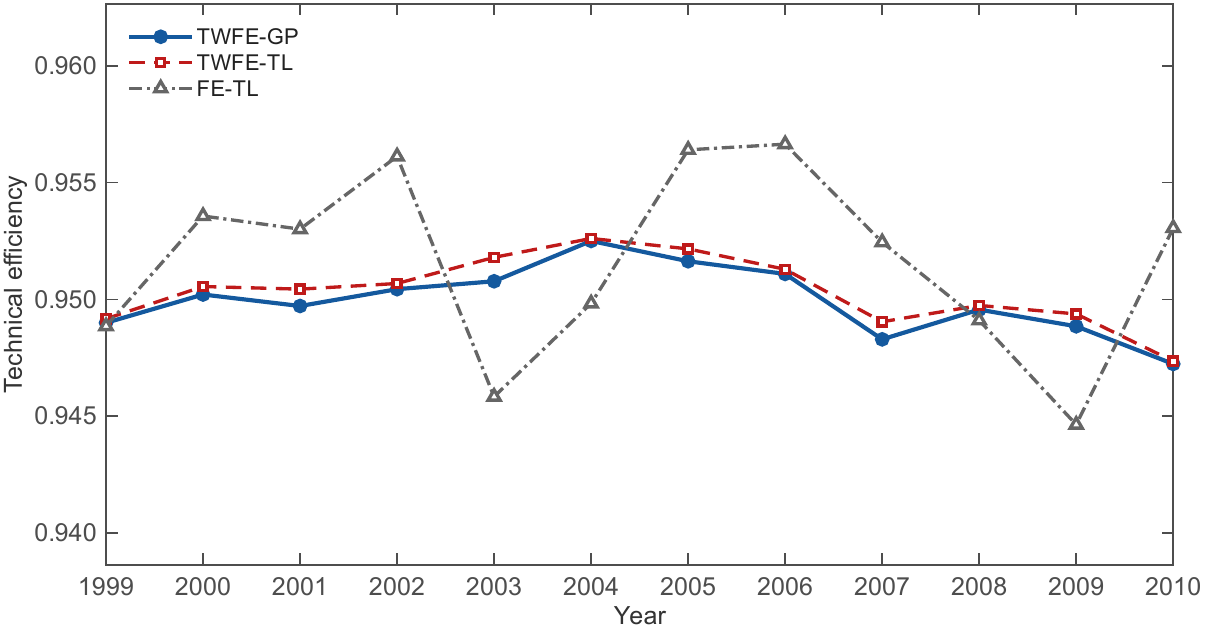}
    \caption{Estimated average technical efficiency over time.}
    \label{fig1}
\end{figure}

Figure~\ref{fig1} plots average technical efficiency by year.
The two models with year effects produce closely aligned profiles,
rising from approximately $0.949$ in 1999 to a peak near $0.953$ in
2004 and ending near $0.947$ in 2010. Their annual averages vary by
approximately half a percentage point over the sample period.
The FE--TL profile varies more, approximately between $0.945$ and
$0.957$, with troughs in 2003 and 2009 and peaks in 2002 and
2005--2006. The figure describes differences between point estimates;
it does not provide uncertainty intervals for these differences.

This pattern is consistent with the models allocating common annual
variation differently between frontier shifts and residual components.
The two-way specifications accommodate unrestricted annual additive
shifts, whereas FE--TL imposes a quadratic trend and allows input
responses to change linearly with time. The comparison demonstrates
sensitivity to the temporal specification, but does not identify the
underlying shocks or establish that the additional variation under
FE--TL represents changes in inefficiency.

\begin{figure}[h]
    \centering
    \includegraphics[width=0.7\textwidth]{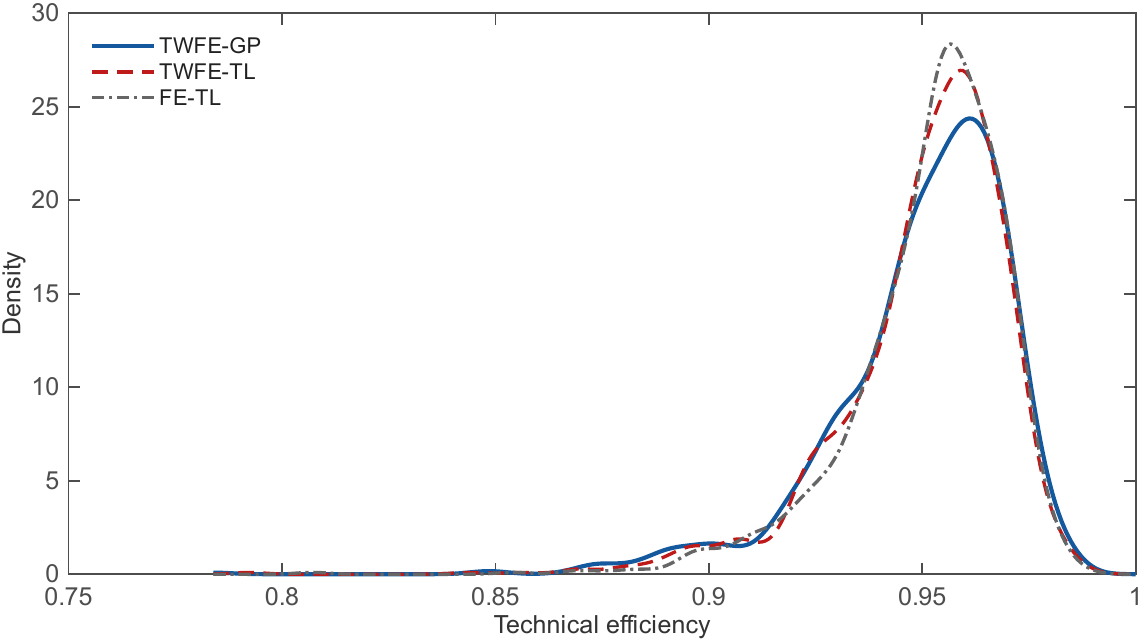}
    \caption{Kernel density estimates of technical efficiency.}
    \label{fig2}
\end{figure}

Figure~\ref{fig2} shows similar distributions of estimated efficiency
scores across observations, with modes near $0.96$ and longer lower
tails. TWFE--GP produces a flatter density than the translog benchmarks,
indicating slightly greater dispersion. These are distributions of
estimated scores, not posterior densities of average efficiency.

\subsection{Input elasticities and returns to scale}
\label{subsec:application_elasticities}

Because output and inputs enter in logarithms, input elasticities are
the derivatives of the log frontier with respect to the log inputs.
For each posterior draw $s$, define the sample-average elasticity of
input $k$ and average RTS by
\[
\bar\varepsilon_k^{(s)}
=\frac{1}{n}\sum_{i=1}^{N}\sum_{t=1}^{T}
   \varepsilon_{it,k}^{(s)},
\qquad
\overline{\mathrm{RTS}}^{(s)}
=\sum_{k=1}^{5}\bar\varepsilon_k^{(s)},
\qquad n=NT.
\]
For TWFE--GP, derivative inference follows
Section~\ref{subsec:derivatives}. Credible intervals for these averages
must account for posterior covariance across both inputs and evaluation
points. For TWFE--TL, the analytical elasticity at each draw is
\[
\varepsilon_{it,k}^{(s)}
=\beta_k^{(s)}+\sum_{j=1}^{5}\beta_{kj}^{(s)}\log x_{j,it}.
\]
For FE--TL, this expression additionally contains
$\beta_{kt}^{(s)}t$. If log inputs are standardized in computation,
the derivatives are converted back to the original log-input scale
using the chain rule in Section~\ref{subsec:derivatives}.

\begin{figure}[h]
    \centering
    \includegraphics[width=0.7\textwidth]{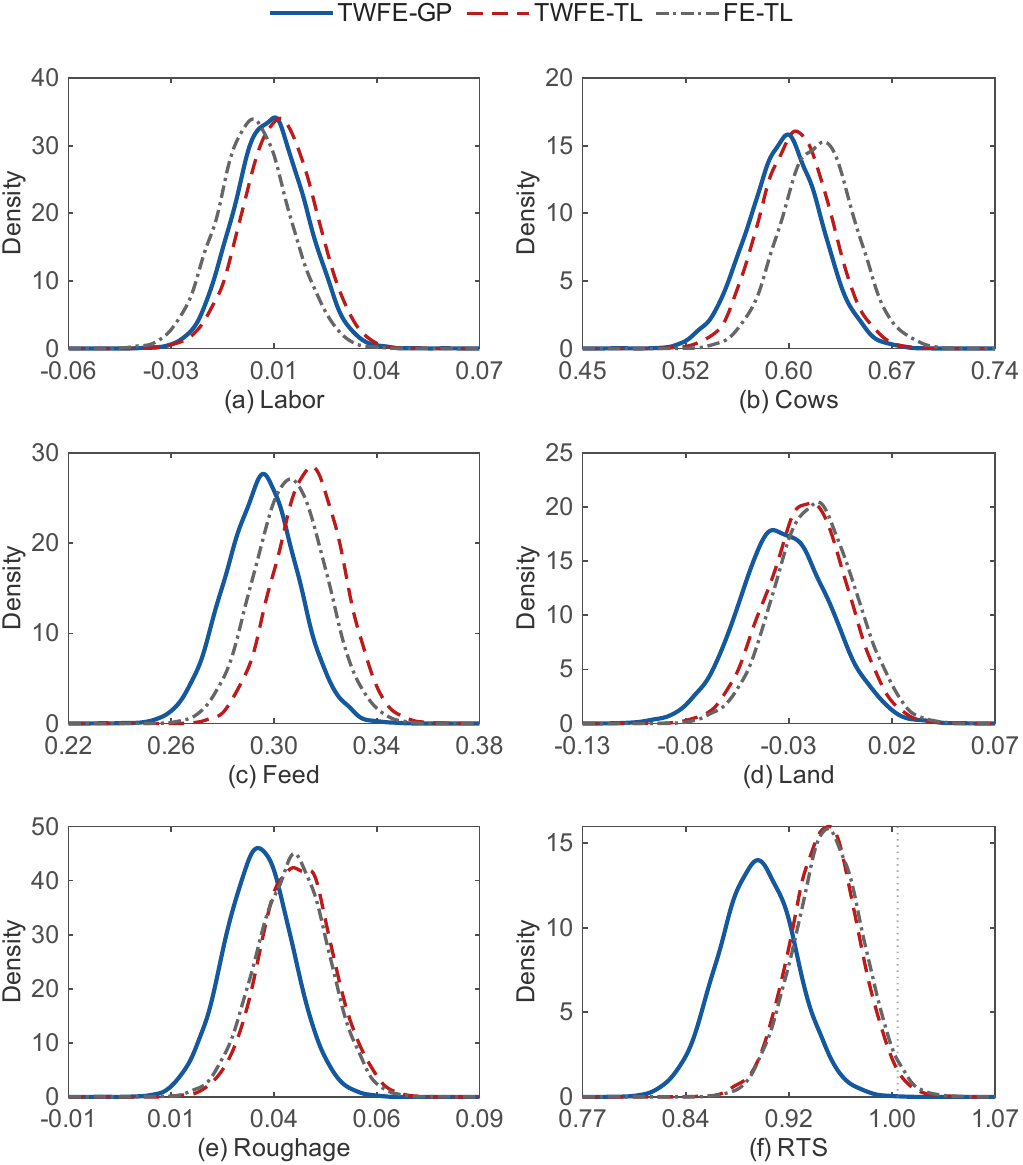}
    \caption{Kernel density estimates of input elasticities and RTS.}
    \label{fig3}
\end{figure}

Table~\ref{estimate} reports posterior summaries of the average
elasticities and RTS, and Figure~\ref{fig3} plots their posterior
density estimates. Cows and purchased feed have the largest estimated
average elasticities. Their posterior means are $0.5937$ and $0.2964$
under TWFE--GP, $0.6014$ and $0.3156$ under TWFE--TL, and $0.6174$
and $0.3079$ under FE--TL. Roughage has a smaller positive average
elasticity, with posterior means of $0.0351$, $0.0447$, and $0.0435$,
respectively. The 95\% credible intervals for these three inputs lie
above zero in all models.

The average labor elasticity is close to zero, and its 95\% credible
interval includes zero under every specification. The average land
elasticity has a negative posterior mean in each model ($-0.0337$ under
TWFE--GP, $-0.0232$ under TWFE--TL, and $-0.0193$ under FE--TL). Its
95\% credible interval also includes zero under every model.
Negative land effects have been reported in earlier analyses of these
data \citep{amsler2016,Amsler2019}, but the present estimates do not
precisely determine the sign of the average land elasticity. They
therefore do not establish that reducing land use would increase
milk output. The GP specification does not impose monotonicity,
and the elasticity estimates are interpreted under the maintained
model assumptions and over the observed input support.

The clearest difference concerns average returns to scale. The reported
posterior mean is $0.8967$ under TWFE--GP, compared with $0.9465$ under
TWFE--TL and $0.9493$ under FE--TL. The corresponding 95\% credible
intervals are $(0.8418,0.9522)$, $(0.8971,0.9947)$, and $(0.8994,0.9989)$.
All three intervals lie below one. The GP estimate has a lower posterior
mean and a wider interval than either translog estimate. Average returns
below one are qualitatively consistent with \citet{parmeter2025},
although their model and estimands differ.

At relative eigenvalue cutoffs of $10^{-9}$, $10^{-10}$, and $10^{-11}$,
the TWFE--GP posterior means of all five sample-average elasticities and
average RTS agree to four decimal places. The largest change in a
reported 95\% credible interval endpoint is $0.0006$, for the lower
bound of the average land elasticity.

Table~\ref{tab:rts_by_herd_size} groups farms into thirds by their 1999
herd size, keeping ties together. For each group, $M_g$ is the mean of its
farm-year RTS values and $H_g$ is the share of its
farm-years with returns below one. Across all farms, the reported posterior
probability that $M_g<1$ is $0.9998$ under TWFE--GP and $0.9855$ under
TWFE--TL. The corresponding posterior means of $H_g$ are $0.6900$ and
$0.8207$. In both models, the 95\% credible intervals for the differences
between the largest and smallest herd-size groups include zero for both
$M_g$ and $H_g$.

\section{Conclusion}
\label{sec:conclude}

This paper develops a Bayesian stochastic frontier model with a Gaussian
process frontier and unrestricted individual and time effects. Orthogonal
contrasts eliminate both sets of fixed-effect coefficients while retaining
the original observation-specific inefficiencies. The resulting restricted
likelihood supports posterior inference for efficiency and frontier
derivatives. Structural identification rests on explicit assumptions;
the GP prior supplies regularization rather than an identification
argument. Simulations show competitive performance under correct
parametric specification and gains when benchmark models restrict frontier
shape incorrectly or omit time effects.

For dairy farms in Asturias, average efficiency is similar across models,
but the GP specification yields a lower posterior mean of average returns
to scale and a wider credible interval. The reported 95\% credible
intervals for average returns lie below one in all three models. Models with year effects produce more stable estimated
efficiency profiles than the benchmark with a quadratic trend and
time--input interactions. The application therefore highlights sensitivity
to both technological and temporal restrictions. Interpretation remains
conditional on the maintained input exogeneity, composed-error distribution,
and identification conditions. The theoretical score correction and
finite-sample evidence do not establish general fixed-$T$ consistency of
the feasible estimator.

\section*{Acknowledgments}
This research was funded by the National Natural Science Foundation of China under Project No.12271047, the Guangdong Provincial Key Laboratory of IRADS (2022B1212010006), Guangdong and Hong Kong Universities “1+1+1” Joint Research Collaboration Scheme (2025A0505000010) and Guangdong Higher Education Upgrading Plan with No. of R0400008-26.

\baselineskip 0.68cm
\setlength{\bibsep}{0pt}
\bibliography{reference}


\begin{table}[!htbp]\centering
\caption{Monte Carlo results for DGP1.}
\label{tab:mc_dgp1}
\begingroup\footnotesize\setlength{\tabcolsep}{3pt}
\begin{tabular}{llrrrrrrrrr}
\toprule
& & \multicolumn{2}{c}{$\sigma_u$} & \multicolumn{2}{c}{$\sigma_v$} & \multicolumn{2}{c}{Frontier} & \multicolumn{2}{c}{Efficiency} & Rank \\
\cmidrule(lr){3-4}\cmidrule(lr){5-6}\cmidrule(lr){7-8}\cmidrule(lr){9-10}
$(N,T)$ & Model & Bias & RMSE & Bias & RMSE & ABIAS & RAMSE & ABIAS & RAMSE & corr. \\
\midrule
$(25, 10)$ & FE--TL & $-0.0224$ & $0.0482$ & $0.0231$ & $0.0351$ & $-0.0190$ & $0.0643$ & $0.0106$ & $0.0860$ & $0.9013$ \\
 & TWFE--TL & $-0.0264$ & $0.0519$ & $0.0282$ & $0.0394$ & $-0.0222$ & $0.0681$ & $0.0124$ & $0.0933$ & $0.8834$ \\
 & FE--GP & $-0.0216$ & $0.0484$ & $0.0236$ & $0.0358$ & $-0.0182$ & $0.0734$ & $0.0102$ & $0.0881$ & $0.8960$ \\
 & TWFE--GP & $-0.0263$ & $0.0532$ & $0.0292$ & $0.0408$ & $-0.0220$ & $0.0775$ & $0.0123$ & $0.0954$ & $0.8782$ \\
\addlinespace
$(25, 20)$ & FE--TL & $-0.0054$ & $0.0266$ & $0.0074$ & $0.0187$ & $-0.0058$ & $0.0405$ & $0.0034$ & $0.0717$ & $0.9294$ \\
 & TWFE--TL & $-0.0077$ & $0.0282$ & $0.0096$ & $0.0213$ & $-0.0078$ & $0.0432$ & $0.0045$ & $0.0793$ & $0.9130$ \\
 & FE--GP & $-0.0049$ & $0.0269$ & $0.0074$ & $0.0189$ & $-0.0052$ & $0.0491$ & $0.0031$ & $0.0732$ & $0.9262$ \\
 & TWFE--GP & $-0.0069$ & $0.0283$ & $0.0097$ & $0.0213$ & $-0.0070$ & $0.0519$ & $0.0041$ & $0.0808$ & $0.9094$ \\
\addlinespace
$(50, 10)$ & FE--TL & $-0.0078$ & $0.0287$ & $0.0106$ & $0.0225$ & $-0.0085$ & $0.0423$ & $0.0046$ & $0.0811$ & $0.9086$ \\
 & TWFE--TL & $-0.0091$ & $0.0292$ & $0.0123$ & $0.0240$ & $-0.0095$ & $0.0432$ & $0.0051$ & $0.0847$ & $0.8999$ \\
 & FE--GP & $-0.0077$ & $0.0294$ & $0.0110$ & $0.0232$ & $-0.0084$ & $0.0503$ & $0.0045$ & $0.0825$ & $0.9053$ \\
 & TWFE--GP & $-0.0087$ & $0.0297$ & $0.0125$ & $0.0244$ & $-0.0092$ & $0.0510$ & $0.0049$ & $0.0860$ & $0.8965$ \\
\addlinespace
$(50, 20)$ & FE--TL & $-0.0027$ & $0.0190$ & $0.0027$ & $0.0128$ & $-0.0028$ & $0.0291$ & $0.0015$ & $0.0705$ & $0.9306$ \\
 & TWFE--TL & $-0.0028$ & $0.0194$ & $0.0028$ & $0.0136$ & $-0.0030$ & $0.0300$ & $0.0016$ & $0.0742$ & $0.9228$ \\
 & FE--GP & $-0.0023$ & $0.0191$ & $0.0026$ & $0.0130$ & $-0.0025$ & $0.0357$ & $0.0013$ & $0.0714$ & $0.9288$ \\
 & TWFE--GP & $-0.0025$ & $0.0194$ & $0.0028$ & $0.0138$ & $-0.0026$ & $0.0365$ & $0.0014$ & $0.0750$ & $0.9210$ \\
\bottomrule
\end{tabular}\endgroup
\end{table}

\begin{table}[tbp]\centering
\caption{Monte Carlo results for DGP2.}
\label{tab:mc_dgp2}
\begingroup\footnotesize\setlength{\tabcolsep}{3pt}
\begin{tabular}{llrrrrrrrrr}
\toprule
& & \multicolumn{2}{c}{$\sigma_u$} & \multicolumn{2}{c}{$\sigma_v$} & \multicolumn{2}{c}{Frontier} & \multicolumn{2}{c}{Efficiency} & Rank \\
\cmidrule(lr){3-4}\cmidrule(lr){5-6}\cmidrule(lr){7-8}\cmidrule(lr){9-10}
$(N,T)$ & Model & Bias & RMSE & Bias & RMSE & ABIAS & RAMSE & ABIAS & RAMSE & corr. \\
\midrule
$(25, 10)$ & FE--TL & $-0.0172$ & $0.0673$ & $0.0806$ & $0.0913$ & $-0.0146$ & $0.1662$ & $0.0092$ & $0.1163$ & $0.8210$ \\
 & TWFE--TL & $-0.0244$ & $0.0744$ & $0.0846$ & $0.0966$ & $-0.0204$ & $0.1688$ & $0.0127$ & $0.1224$ & $0.8038$ \\
 & FE--GP & $-0.0223$ & $0.0493$ & $0.0288$ & $0.0402$ & $-0.0203$ & $0.0939$ & $0.0114$ & $0.0932$ & $0.8834$ \\
 & TWFE--GP & $-0.0286$ & $0.0552$ & $0.0352$ & $0.0472$ & $-0.0253$ & $0.0990$ & $0.0144$ & $0.1008$ & $0.8646$ \\
\addlinespace
$(25, 20)$ & FE--TL & $0.0027$ & $0.0361$ & $0.0696$ & $0.0743$ & $0.0027$ & $0.1568$ & $-0.0012$ & $0.1046$ & $0.8461$ \\
 & TWFE--TL & $0.0006$ & $0.0394$ & $0.0705$ & $0.0762$ & $0.0010$ & $0.1573$ & $-0.0002$ & $0.1098$ & $0.8297$ \\
 & FE--GP & $-0.0080$ & $0.0268$ & $0.0103$ & $0.0207$ & $-0.0068$ & $0.0660$ & $0.0036$ & $0.0770$ & $0.9178$ \\
 & TWFE--GP & $-0.0103$ & $0.0288$ & $0.0128$ & $0.0242$ & $-0.0087$ & $0.0688$ & $0.0047$ & $0.0844$ & $0.9006$ \\
\addlinespace
$(50, 10)$ & FE--TL & $0.0022$ & $0.0441$ & $0.0709$ & $0.0771$ & $0.0023$ & $0.1558$ & $-0.0005$ & $0.1112$ & $0.8268$ \\
 & TWFE--TL & $0.0017$ & $0.0446$ & $0.0709$ & $0.0776$ & $0.0019$ & $0.1560$ & $-0.0002$ & $0.1136$ & $0.8185$ \\
 & FE--GP & $-0.0102$ & $0.0317$ & $0.0152$ & $0.0262$ & $-0.0086$ & $0.0665$ & $0.0050$ & $0.0862$ & $0.8965$ \\
 & TWFE--GP & $-0.0113$ & $0.0321$ & $0.0166$ & $0.0268$ & $-0.0095$ & $0.0678$ & $0.0055$ & $0.0898$ & $0.8872$ \\
\addlinespace
$(50, 20)$ & FE--TL & $0.0061$ & $0.0258$ & $0.0696$ & $0.0719$ & $0.0054$ & $0.1538$ & $-0.0026$ & $0.1032$ & $0.8484$ \\
 & TWFE--TL & $0.0057$ & $0.0264$ & $0.0698$ & $0.0724$ & $0.0050$ & $0.1538$ & $-0.0024$ & $0.1058$ & $0.8402$ \\
 & FE--GP & $-0.0039$ & $0.0185$ & $0.0053$ & $0.0144$ & $-0.0031$ & $0.0491$ & $0.0017$ & $0.0738$ & $0.9240$ \\
 & TWFE--GP & $-0.0041$ & $0.0186$ & $0.0056$ & $0.0150$ & $-0.0033$ & $0.0498$ & $0.0018$ & $0.0774$ & $0.9159$ \\
\bottomrule
\end{tabular}\endgroup
\end{table}

\begin{table}[tbp]\centering
\caption{Monte Carlo results for DGP3.}
\label{tab:mc_dgp3}
\begingroup\footnotesize\setlength{\tabcolsep}{3pt}
\begin{tabular}{llrrrrrrrrr}
\toprule
& & \multicolumn{2}{c}{$\sigma_u$} & \multicolumn{2}{c}{$\sigma_v$} & \multicolumn{2}{c}{Frontier} & \multicolumn{2}{c}{Efficiency} & Rank \\
\cmidrule(lr){3-4}\cmidrule(lr){5-6}\cmidrule(lr){7-8}\cmidrule(lr){9-10}
$(N,T)$ & Model & Bias & RMSE & Bias & RMSE & ABIAS & RAMSE & ABIAS & RAMSE & corr. \\
\midrule
$(25, 10)$ & FE--TL & $-0.0118$ & $0.0706$ & $0.0925$ & $0.1022$ & $-0.0094$ & $0.2671$ & $0.0064$ & $0.1212$ & $0.8048$ \\
 & TWFE--TL & $-0.0215$ & $0.0484$ & $0.0262$ & $0.0371$ & $-0.0189$ & $0.0763$ & $0.0106$ & $0.0923$ & $0.8855$ \\
 & FE--GP & $-0.0096$ & $0.0705$ & $0.0909$ & $0.1008$ & $-0.0074$ & $0.2573$ & $0.0054$ & $0.1221$ & $0.8010$ \\
 & TWFE--GP & $-0.0209$ & $0.0498$ & $0.0271$ & $0.0392$ & $-0.0184$ & $0.0910$ & $0.0104$ & $0.0949$ & $0.8791$ \\
\addlinespace
$(25, 20)$ & FE--TL & $-0.0090$ & $0.0420$ & $0.0940$ & $0.0975$ & $-0.0066$ & $0.2591$ & $0.0039$ & $0.1087$ & $0.8327$ \\
 & TWFE--TL & $-0.0099$ & $0.0294$ & $0.0108$ & $0.0212$ & $-0.0081$ & $0.0487$ & $0.0046$ & $0.0794$ & $0.9128$ \\
 & FE--GP & $-0.0084$ & $0.0419$ & $0.0939$ & $0.0975$ & $-0.0061$ & $0.2532$ & $0.0037$ & $0.1094$ & $0.8301$ \\
 & TWFE--GP & $-0.0098$ & $0.0293$ & $0.0113$ & $0.0217$ & $-0.0080$ & $0.0583$ & $0.0045$ & $0.0809$ & $0.9093$ \\
\addlinespace
$(50, 10)$ & FE--TL & $-0.0116$ & $0.0427$ & $0.0808$ & $0.0862$ & $-0.0099$ & $0.1779$ & $0.0056$ & $0.1093$ & $0.8320$ \\
 & TWFE--TL & $-0.0091$ & $0.0287$ & $0.0123$ & $0.0244$ & $-0.0090$ & $0.0481$ & $0.0050$ & $0.0846$ & $0.9005$ \\
 & FE--GP & $-0.0116$ & $0.0429$ & $0.0806$ & $0.0862$ & $-0.0100$ & $0.1738$ & $0.0056$ & $0.1099$ & $0.8302$ \\
 & TWFE--GP & $-0.0094$ & $0.0292$ & $0.0131$ & $0.0249$ & $-0.0091$ & $0.0545$ & $0.0051$ & $0.0857$ & $0.8978$ \\
\addlinespace
$(50, 20)$ & FE--TL & $-0.0029$ & $0.0294$ & $0.1029$ & $0.1046$ & $-0.0027$ & $0.3376$ & $0.0016$ & $0.1109$ & $0.8218$ \\
 & TWFE--TL & $-0.0029$ & $0.0195$ & $0.0035$ & $0.0141$ & $-0.0031$ & $0.0351$ & $0.0017$ & $0.0742$ & $0.9232$ \\
 & FE--GP & $-0.0024$ & $0.0294$ & $0.1025$ & $0.1041$ & $-0.0022$ & $0.3352$ & $0.0013$ & $0.1113$ & $0.8206$ \\
 & TWFE--GP & $-0.0021$ & $0.0197$ & $0.0029$ & $0.0144$ & $-0.0023$ & $0.0419$ & $0.0013$ & $0.0752$ & $0.9212$ \\
\bottomrule
\end{tabular}\endgroup
\end{table}

\begin{table}[tbp]\centering
\caption{Monte Carlo results for DGP4.}
\label{tab:mc_dgp4}
\begingroup\footnotesize\setlength{\tabcolsep}{3pt}
\begin{tabular}{llrrrrrrrrr}
\toprule
& & \multicolumn{2}{c}{$\sigma_u$} & \multicolumn{2}{c}{$\sigma_v$} & \multicolumn{2}{c}{Frontier} & \multicolumn{2}{c}{Efficiency} & Rank \\
\cmidrule(lr){3-4}\cmidrule(lr){5-6}\cmidrule(lr){7-8}\cmidrule(lr){9-10}
$(N,T)$ & Model & Bias & RMSE & Bias & RMSE & ABIAS & RAMSE & ABIAS & RAMSE & corr. \\
\midrule
$(25, 10)$ & FE--TL & $-0.0127$ & $0.0996$ & $0.1336$ & $0.1440$ & $-0.0089$ & $0.3114$ & $0.0084$ & $0.1393$ & $0.7513$ \\
 & TWFE--TL & $-0.0140$ & $0.0697$ & $0.0772$ & $0.0879$ & $-0.0107$ & $0.1706$ & $0.0074$ & $0.1194$ & $0.8083$ \\
 & FE--GP & $-0.0127$ & $0.0796$ & $0.0957$ & $0.1064$ & $-0.0093$ & $0.2650$ & $0.0071$ & $0.1261$ & $0.7899$ \\
 & TWFE--GP & $-0.0252$ & $0.0534$ & $0.0332$ & $0.0447$ & $-0.0211$ & $0.1108$ & $0.0122$ & $0.1000$ & $0.8652$ \\
\addlinespace
$(25, 20)$ & FE--TL & $-0.0064$ & $0.0529$ & $0.1412$ & $0.1445$ & $-0.0057$ & $0.3014$ & $0.0037$ & $0.1250$ & $0.7777$ \\
 & TWFE--TL & $0.0015$ & $0.0388$ & $0.0712$ & $0.0763$ & $0.0006$ & $0.1562$ & $0.0000$ & $0.1082$ & $0.8346$ \\
 & FE--GP & $-0.0068$ & $0.0406$ & $0.0975$ & $0.1010$ & $-0.0061$ & $0.2568$ & $0.0034$ & $0.1112$ & $0.8237$ \\
 & TWFE--GP & $-0.0073$ & $0.0286$ & $0.0129$ & $0.0236$ & $-0.0074$ & $0.0729$ & $0.0040$ & $0.0840$ & $0.9016$ \\
\addlinespace
$(50, 10)$ & FE--TL & $-0.0128$ & $0.0565$ & $0.1250$ & $0.1298$ & $-0.0105$ & $0.2256$ & $0.0063$ & $0.1252$ & $0.7783$ \\
 & TWFE--TL & $-0.0010$ & $0.0420$ & $0.0629$ & $0.0701$ & $-0.0011$ & $0.1443$ & $0.0010$ & $0.1095$ & $0.8308$ \\
 & FE--GP & $-0.0114$ & $0.0462$ & $0.0825$ & $0.0884$ & $-0.0095$ & $0.1761$ & $0.0054$ & $0.1123$ & $0.8215$ \\
 & TWFE--GP & $-0.0108$ & $0.0328$ & $0.0161$ & $0.0270$ & $-0.0100$ & $0.0689$ & $0.0055$ & $0.0890$ & $0.8892$ \\
\addlinespace
$(50, 20)$ & FE--TL & $-0.0035$ & $0.0374$ & $0.1581$ & $0.1596$ & $-0.0030$ & $0.3806$ & $0.0020$ & $0.1281$ & $0.7582$ \\
 & TWFE--TL & $0.0043$ & $0.0293$ & $0.0768$ & $0.0793$ & $0.0034$ & $0.1613$ & $-0.0015$ & $0.1072$ & $0.8353$ \\
 & FE--GP & $-0.0047$ & $0.0302$ & $0.1056$ & $0.1073$ & $-0.0039$ & $0.3374$ & $0.0023$ & $0.1126$ & $0.8153$ \\
 & TWFE--GP & $-0.0034$ & $0.0201$ & $0.0065$ & $0.0163$ & $-0.0034$ & $0.0544$ & $0.0020$ & $0.0775$ & $0.9159$ \\
\bottomrule
\end{tabular}\endgroup
\end{table}

\begin{table}[tbp]\centering
\caption{Estimation under DGP1 using dummy-variable maximum likelihood and the invariant likelihood.}
\label{tab:tfe}
\begingroup\footnotesize\setlength{\tabcolsep}{3pt}
\begin{tabular}{llrrrrrrrrr}
\toprule
& & \multicolumn{2}{c}{$\sigma_u$} & \multicolumn{2}{c}{$\sigma_v$} & \multicolumn{2}{c}{Frontier} & \multicolumn{2}{c}{Efficiency} & Rank \\
\cmidrule(lr){3-4}\cmidrule(lr){5-6}\cmidrule(lr){7-8}\cmidrule(lr){9-10}
$(N,T)$ & Model & Bias & RMSE & Bias & RMSE & ABIAS & RAMSE & ABIAS & RAMSE & corr. \\
\midrule
$(25, 10)$ & TFE--MLE & $-0.0225$ & $0.0363$ & $-0.0994$ & $0.0998$ & $-0.0353$ & $0.0713$ & $0.0260$ & $0.0997$ & $0.8890$ \\
 & FE--TL & $-0.0224$ & $0.0482$ & $0.0231$ & $0.0351$ & $-0.0190$ & $0.0643$ & $0.0106$ & $0.0860$ & $0.9013$ \\
\addlinespace
$(25, 20)$ & TFE--MLE & $0.0185$ & $0.0332$ & $-0.0783$ & $0.0871$ & $0.0144$ & $0.0491$ & $-0.0066$ & $0.0808$ & $0.9205$ \\
 & FE--TL & $-0.0054$ & $0.0266$ & $0.0074$ & $0.0187$ & $-0.0058$ & $0.0405$ & $0.0034$ & $0.0717$ & $0.9294$ \\
\addlinespace
$(50, 10)$ & TFE--MLE & $-0.0143$ & $0.0245$ & $-0.1000$ & $0.1000$ & $-0.0298$ & $0.0518$ & $0.0225$ & $0.0955$ & $0.8972$ \\
 & FE--TL & $-0.0078$ & $0.0287$ & $0.0106$ & $0.0225$ & $-0.0085$ & $0.0423$ & $0.0046$ & $0.0811$ & $0.9086$ \\
\addlinespace
$(50, 20)$ & TFE--MLE & $0.0225$ & $0.0313$ & $-0.0802$ & $0.0878$ & $0.0190$ & $0.0393$ & $-0.0097$ & $0.0802$ & $0.9219$ \\
 & FE--TL & $-0.0027$ & $0.0190$ & $0.0027$ & $0.0128$ & $-0.0028$ & $0.0291$ & $0.0015$ & $0.0705$ & $0.9306$ \\
\bottomrule
\end{tabular}\endgroup
\end{table}

\begin{table}[!hb]\centering
\caption{Descriptive statistics for 1,080 farm-years in Asturias, 1999--2010.}
\label{table:descriptivestats}
\begin{tabular}{lrrrr}
\toprule
Variable & Mean & SD & Minimum & Maximum \\
\midrule
Annual milk output (liters) & 372,323 & 209,447 & 43,090 & 1,322,276 \\
Labor (full-time equivalents) & 1.94 & 0.77 & 0.37 & 5.74 \\
Milking cows (head) & 46.29 & 21.21 & 11 & 134 \\
Purchased dairy feed (kg) & 170,650 & 101,542 & 16,982 & 652,574 \\
Agricultural land (ha) & 20.23 & 9.19 & 4.65 & 56.58 \\
Forage expenses (2010 euros) & 24,566.79 & 17,402.10 & 133.08 & 132,750.03 \\
\bottomrule
\end{tabular}
\end{table}

\begin{table}[tbp]\centering
\caption{Posterior estimates for the dairy production frontiers.}
\label{estimate}
\begingroup\small\setlength{\tabcolsep}{3.0pt}
\begin{tabular}{lrlrlrl}
\toprule
& \multicolumn{2}{c}{TWFE--GP} & \multicolumn{2}{c}{TWFE--TL} & \multicolumn{2}{c}{FE--TL} \\
\cmidrule(lr){2-3}\cmidrule(lr){4-5}\cmidrule(lr){6-7}
Quantity & Mean & 95\% interval & Mean & 95\% interval & Mean & 95\% interval \\
\midrule
$\sigma_u$ & $0.0657$ & $(0.0486,0.0807)$ & $0.0651$ & $(0.0466,0.0826)$ & $0.0634$ & $(0.0450,0.0808)$ \\
$\sigma_v$ & $0.0545$ & $(0.0473,0.0617)$ & $0.0605$ & $(0.0531,0.0673)$ & $0.0627$ & $(0.0556,0.0691)$ \\
\addlinespace
$\sigma_f$ & $0.5086$ & $(0.3910,0.6593)$ & -- & -- & -- & -- \\
$\log\ell_1$ & $2.3928$ & $(1.4688,3.7918)$ & -- & -- & -- & -- \\
$\log\ell_2$ & $0.7415$ & $(0.3317,1.1318)$ & -- & -- & -- & -- \\
$\log\ell_3$ & $0.9125$ & $(0.5318,1.2785)$ & -- & -- & -- & -- \\
$\log\ell_4$ & $0.5680$ & $(0.1643,1.0072)$ & -- & -- & -- & -- \\
$\log\ell_5$ & $1.8204$ & $(1.3363,2.3514)$ & -- & -- & -- & -- \\
\addlinespace
WAIC & $-2588.5$ & -- & $-2484.4$ & -- & $-2449.7$ & -- \\
\addlinespace
\multicolumn{7}{l}{Sample-average elasticities and returns to scale} \\
Labor & $0.0053$ & $(-0.0165,0.0277)$ & $0.0080$ & $(-0.0148,0.0304)$ & $-0.0002$ & $(-0.0234,0.0233)$ \\
Cows & $0.5937$ & $(0.5417,0.6445)$ & $0.6014$ & $(0.5544,0.6503)$ & $0.6174$ & $(0.5677,0.6677)$ \\
Feed & $0.2964$ & $(0.2673,0.3254)$ & $0.3156$ & $(0.2884,0.3424)$ & $0.3079$ & $(0.2794,0.3366)$ \\
Land & $-0.0337$ & $(-0.0765,0.0099)$ & $-0.0232$ & $(-0.061,0.0135)$ & $-0.0193$ & $(-0.0574,0.0181)$ \\
Roughage & $0.0351$ & $(0.0181,0.0522)$ & $0.0447$ & $(0.0273,0.0623)$ & $0.0435$ & $(0.0257,0.0607)$ \\
Average RTS & $0.8967$ & $(0.8418,0.9522)$ & $0.9465$ & $(0.8971,0.9947)$ & $0.9493$ & $(0.8994,0.9989)$ \\
\addlinespace
\multicolumn{7}{l}{Technical efficiency across farm-years} \\
Mean score & $0.9499$ & -- & $0.9504$ & -- & $0.9516$ & -- \\
Minimum score & $0.7839$ & -- & $0.7921$ & -- & $0.8071$ & -- \\
Maximum score & $0.9864$ & -- & $0.9876$ & -- & $0.9880$ & -- \\
\bottomrule
\end{tabular}\endgroup
\par\smallskip\begin{minipage}{0.94\linewidth}\footnotesize Intervals are 95\% posterior credible intervals. Efficiency scores are posterior means for individual farm-years. WAIC values use the deviance scale and are reported as model-specific diagnostics.\end{minipage}
\end{table}

\begin{table}[tbp]\centering
\caption{Returns to scale by initial herd size.}
\label{tab:rts_by_herd_size}
\begingroup\small\setlength{\tabcolsep}{4.15pt}
\begin{tabular}{@{}llrcrc@{}}
\toprule
Model & Herd size & Farms & $M_g$ & $\Pr(M_g<1)$ & $H_g$ \\
\midrule
TWFE--GP & All farms & 90 & $0.8967\;(0.8418,0.9522)$ & $0.9998$ & $0.6900\;(0.5574,0.8324)$ \\
 & Small & 30 & $0.8866\;(0.8099,0.9627)$ & $0.9978$ & $0.6285\;(0.4500,0.8361)$ \\
 & Medium & 30 & $0.9044\;(0.8353,0.9760)$ & $0.9956$ & $0.7259\;(0.5250,0.9222)$ \\
 & Large & 30 & $0.8991\;(0.8287,0.9689)$ & $0.9971$ & $0.7156\;(0.5944,0.8278)$ \\
\addlinespace[2pt]
 & Large $-$ small & -- & $0.0125\;(-0.0831,0.1060)$ & -- & $0.0871\;(-0.1139,0.2778)$ \\
\midrule
TWFE--TL & All farms & 90 & $0.9465\;(0.8971,0.9947)$ & $0.9855$ & $0.8207\;(0.5648,0.9750)$ \\
 & Small & 30 & $0.9591\;(0.8943,1.0229)$ & $0.8992$ & $0.7480\;(0.3694,0.9611)$ \\
 & Medium & 30 & $0.9492\;(0.9004,0.9970)$ & $0.9820$ & $0.8340\;(0.5250,0.9944)$ \\
 & Large & 30 & $0.9310\;(0.8724,0.9884)$ & $0.9915$ & $0.8800\;(0.6333,0.9861)$ \\
\addlinespace[2pt]
 & Large $-$ small & -- & $-0.0281\;(-0.1011,0.0448)$ & -- & $0.1320\;(-0.1458,0.4972)$ \\
\bottomrule
\end{tabular}\endgroup
\par\smallskip\begin{minipage}{0.94\linewidth}\footnotesize Herd-size groups are thirds of the 1999 herd-size distribution, with ties kept together. $M_g$ is mean returns to scale in group $g$, and $H_g$ is the share of its farm-years with returns below one. Entries for $M_g$ and $H_g$ are posterior means with 95\% credible intervals. The last row for each model is the draw-by-draw difference between the large and small groups.\end{minipage}
\end{table}


\clearpage 

\pagebreak 

\baselineskip 0.78cm

\setcounter{page}{1}

\noindent{\large Supplementary Material}
\appendix 
\numberwithin{equation}{subsection}
\counterwithin{table}{section}

\section{Proofs for Section~\ref{sec:model}}
\label{app:section2_proofs}

Throughout this appendix, $D$ is a fixed $n\times q$ matrix of rank
$q<n$, $r=n-q$, and $R$ is an orthonormal basis of
$\mathcal C(D)^\perp$. All probability statements involving the error
components condition on the complete input array $X$. Structural
identification treats $f$ as fixed; Gaussian covariance-score calculations
instead integrate $f$ under its stated GP prior.

\subsection{Proof of Proposition~\ref{prop:contrast_geometry}}
\label{app:proof_contrast_geometry}

\noindent\textbf{Step 1 (Compute the nuisance Gram matrix).}
Let $D_1=C_N\otimes\mathbf1_T$ and
$D_2=\mathbf1_N\otimes C_T$. The Kronecker product identity
$(A\otimes B)'(C\otimes E)=(A'C)\otimes(B'E)$ gives
\[
D_1'D_1=TI_{N-1},\qquad D_2'D_2=NI_{T-1},
\qquad
D_1'D_2=(C_N'\mathbf1_N)\otimes(\mathbf1_T'C_T)=0.
\]
Hence \eqref{eq:DTD} holds and $D'D$ is positive definite, so
$\operatorname{rank}(D)=q=N+T-2$. Also $D'\mathbf1_n=0$.

\medskip
\noindent\textbf{Step 2 (Verify the contrast basis).}
For $Q=C_N\otimes C_T$,
\[
Q'Q=I_{(N-1)(T-1)},\qquad q_0'q_0=1,\qquad q_0'Q=0.
\]
Furthermore,
\[
Q'D_1=I_{N-1}\otimes(C_T'\mathbf1_T)=0,
\qquad
Q'D_2=(C_N'\mathbf1_N)\otimes I_{T-1}=0,
\]
and $q_0'D=0$. Thus $R=[q_0,Q]$ has orthonormal columns in
$\mathcal C(D)^\perp$. Its number of columns is
$1+(N-1)(T-1)=n-q$, so it spans that entire complement. The orthogonal
projector onto this complement is both $RR'$ and
$I_n-D(D'D)^{-1}D'$. This proves \eqref{eq:Rproperties}.

\medskip
\noindent\textbf{Step 3 (Establish maximal invariance).}
For any $b\in\mathbb R^q$, $R'(y+Db)=R'y$. Conversely, if
$R'y_1=R'y_2$, then $R'(y_1-y_2)=0$. Since
$\ker(R')=\mathcal C(D)$, there is a $b$ with $y_1-y_2=Db$.
Thus equal contrasts characterise exactly the translation orbits.
The direction $q_0$ is orthogonal to $D$ by the two zero-sum
normalizations, which explains why it remains in $R$.

\subsection{Support arguments and the primitive identification condition}
\label{app:proof_primitive_identification}

\subsubsection{A lemma}

\begin{lemma}[Extending an almost-sure equality to a support]
\label{lem:support_extension}
Let $Z$ be a random vector with topological support $S$, and let $g$ be
continuous on $S$ in its relative topology. If $g(Z)=0$ almost surely,
then $g(z)=0$ at every $z\in S$.
\end{lemma}

\noindent{\em Proof}. 
\noindent\textbf{Step 1 (Suppose the equality fails at a support point).}
If $g(z_0)\ne0$ for $z_0\in S$, continuity gives a relative
neighborhood $B(z_0,\varepsilon)\cap S$ on which
$|g(z)|>|g(z_0)|/2$.

\medskip
\noindent\textbf{Step 2 (Use the defining property of support).}
Every open ball centered at $z_0\in S$ has positive probability.
Since $Z\in S$ almost surely, it follows that
$P\{|g(Z)|>|g(z_0)|/2\}>0$, contradicting $g(Z)=0$ almost surely.

\subsubsection{Proof of Proposition~\ref{prop:primitive_identification}}

\noindent\textbf{Step 1 (Obtain rectangle identities without fixing random coefficients).}
Take $h\in\mathcal F-\mathcal F$ with $M_Dh_X=0$ almost surely.
For each realization outside a null set, $h_X\in\mathcal C(D)$.
Every vector in this column space has an additive row-and-column
representation. Consequently, for every rectangle,
\begin{equation}
\label{eq:rectangle_identity_appendix}
h(X_{it})-h(X_{is})-h(X_{jt})+h(X_{js})=0
\quad\text{almost surely}.
\end{equation}
There are finitely many rectangles, so all required identities hold
simultaneously outside one null set. Their validity does not require
the row-and-column coefficients representing $h_X$ to be nonrandom.

\medskip
\noindent\textbf{Step 2 (Extend each identity to a product support).}
For a selected rectangle define
\[
g(x_1,x_2,x_3,x_4)=h(x_1)-h(x_2)-h(x_3)+h(x_4).
\]
This is continuous on the product in
\eqref{eq:rectangle_support}. By that support condition,
\eqref{eq:rectangle_identity_appendix}, and
Lemma~\ref{lem:support_extension},
\begin{equation}
\label{eq:rectangle_pointwise_appendix}
h(x_1)-h(x_2)-h(x_3)+h(x_4)=0
\end{equation}
for every point in that product.

\medskip
\noindent\textbf{Step 3 (Vary one coordinate at a time).}
Fix any $x_2\in\mathcal S_{is}$, $x_3\in\mathcal S_{jt}$, and
$x_4\in\mathcal S_{js}$. For arbitrary
$x_1,\widetilde x_1\in\mathcal S_{it}$, subtract the corresponding
versions of \eqref{eq:rectangle_pointwise_appendix}. This gives
$h(x_1)=h(\widetilde x_1)$, so $h$ is constant on
$\mathcal S_{it}$. Repeating this operation for coordinates two,
three, and four shows constancy on every support in that rectangle.
The selected rectangles cover all cells. Thus there are constants
$c_{it}$ such that $h(x)=c_{it}$ on $\mathcal S_{it}$.

\medskip
\noindent\textbf{Step 4 (Propagate equality through overlap).}
If $\mathcal S_{it}\cap\mathcal S_{js}\ne\varnothing$, evaluation
at any common point gives $c_{it}=c_{js}$. Connectedness of the
finite graph then gives one constant $c$ on all of $\mathcal S$.

\medskip
\noindent\textbf{Step 5 (Use the zero-sum nuisance space).}
We now have $h_X=c\mathbf1_n$ almost surely and $h_X\in\mathcal C(D)$
almost surely. But $D'\mathbf1_n=0$ implies
$\mathcal C(D)\perp\mathbf1_n$. Taking the inner product with
$\mathbf1_n$ therefore gives $nc=0$, so $c=0$. This proves the
pointwise conclusion and hence Assumption~1.

\subsubsection{Why marginal support overlap is insufficient}
\label{app:primitive_counterexample}

For $N=T=2$, let $Z\sim\operatorname{Unif}[-1,1]$ and set
\[
X_{11}=X_{12}=Z,\qquad X_{21}=X_{22}=-Z.
\]
All four marginal densities are positive on $(-1,1)$, all supports are
$[-1,1]$, and the overlap graph is complete. Nevertheless, for
$h(x)=x$,
\[
h_X=(Z,Z,-Z,-Z)'\in\mathcal C(D),\qquad M_Dh_X=0
\quad\text{almost surely},
\]
although $h$ is not zero or constant on the support. The row effects
$(Z,-Z)$ satisfy the zero-sum restriction for every realization.
The joint support is restricted to
$\{(z,z,-z,-z):z\in[-1,1]\}$, so
\eqref{eq:rectangle_support} fails.

This does not contradict the valid statement that, for fixed nonrandom
constants $a_i,b_t$, $h(X_{it})=a_i+b_t$ almost surely and continuity
force $h$ to be constant on each cell support. That statement follows
from Lemma~\ref{lem:support_extension}, and overlap equates its
constants. The missing implication would be to replace the
realization-dependent representation furnished by
$M_Dh_X=0$ with fixed constants. The joint-support argument in
Proposition~\ref{prop:primitive_identification} avoids that step.

\subsection{Lemmas and Proof of Theorem~\ref{thm:identification}}
\label{app:proof_identification}
\label{app:proof_identification_steps}

\subsubsection{Lemmas}

\begin{lemma}[Half-normal moments and cumulants]
\label{lem:half_normal_cumulants}
If $U=\sigma_u|Z|$, $Z\sim N(0,1)$, and $\sigma_u\geq0$, then
\begin{align}
&E(U)=\sigma_u\sqrt{2/\pi},
\label{eq:EU_appendix}\\
&\operatorname{Var}(U)=(1-2/\pi)\sigma_u^2,
\label{eq:VarU_appendix}\\
&\kappa_4(U)=8(\pi-3)\sigma_u^4/\pi^2.
\label{eq:kappa4U_appendix}
\end{align}
\end{lemma}

\noindent{\em Proof.}
\noindent\textbf{Step 1 (Compute raw moments).}
For $\sigma_u>0$ and an integer $m\geq0$, substituting
$s=z^2/2$ into the half-normal integral gives
\begin{align}
E(U^m)
&=\sigma_u^m\sqrt{\frac2\pi}
  \int_0^\infty z^m e^{-z^2/2}\,dz
\nonumber\\
&=\sigma_u^m\frac{2^{m/2}}{\sqrt\pi}
  \int_0^\infty s^{(m-1)/2}e^{-s}\,ds
=\sigma_u^m2^{m/2}
  \frac{\Gamma\{(m+1)/2\}}{\sqrt\pi}.
\label{eq:half_normal_raw_moments}
\end{align}
In particular, with $c=\sqrt{2/\pi}$,
\[
E(U)=c\sigma_u,\quad E(U^2)=\sigma_u^2,
\quad E(U^3)=2c\sigma_u^3,\quad E(U^4)=3\sigma_u^4.
\]

\medskip
\noindent\textbf{Step 2 (Compute the variance).}
Subtracting $\{E(U)\}^2$ from $E(U^2)$ gives
$\operatorname{Var}(U)=(1-c^2)\sigma_u^2$, which is
\eqref{eq:VarU_appendix}.

\medskip
\noindent\textbf{Step 3 (Compute the fourth cumulant).}
For $\mu=E(U)$, the fourth central moment is
\begin{align*}
E\{(U-\mu)^4\}
&=E(U^4)-4\mu E(U^3)+6\mu^2E(U^2)-3\mu^4\\
&=\sigma_u^4(3-2c^2-3c^4)
=\sigma_u^4\left(3-\frac4\pi-\frac{12}{\pi^2}\right).
\end{align*}
Subtracting three times the squared variance gives
\[
\kappa_4(U)
=\sigma_u^4\left\{3-\frac4\pi-\frac{12}{\pi^2}
                  -3\left(1-\frac2\pi\right)^2\right\}
=\frac{8(\pi-3)}{\pi^2}\sigma_u^4.
\]
At $\sigma_u=0$, $U=0$ almost surely and all three identities hold
directly.

\begin{lemma}[Cumulants of error contrasts]
\label{lem:contrast_cumulants}
Let $\varepsilon_j=V_j-U_j$ be independent across $j$, with the
normal--half-normal specification in \eqref{eq:u_dist}. For a
deterministic vector $a\in\mathbb R^n$,
\begin{align}
\kappa_2(a'\varepsilon)
&=\{\sigma_v^2+(1-2/\pi)\sigma_u^2\}\sum_j a_j^2,
\label{eq:kappa2_contrast_appendix}\\
\kappa_4(a'\varepsilon)
&=\frac{8(\pi-3)}{\pi^2}\sigma_u^4\sum_j a_j^4.
\label{eq:kappa4_contrast_appendix}
\end{align}
\end{lemma}

\noindent{\em Proof.}
\noindent\textbf{Step 1 (Calculate a single composed error's cumulants).}
Independence gives
$\kappa_2(V_j-U_j)=\sigma_v^2+(1-2/\pi)\sigma_u^2$.
The Gaussian fourth cumulant is zero, and sign reversal leaves an
even-order cumulant unchanged. Thus
$\kappa_4(V_j-U_j)=\kappa_4(U_j)$, given by
Lemma~\ref{lem:half_normal_cumulants}.

\medskip
\noindent\textbf{Step 2 (Add independent contributions).}
Independence makes the cumulant generating function of
$\sum_j a_j\varepsilon_j$ the sum of the individual cumulant
generating functions evaluated at $a_jt$. Differentiating $m$ times at
zero gives
$\kappa_m(\sum_j a_j\varepsilon_j)=\sum_j a_j^m\kappa_m(\varepsilon_j)$.
Apply this identity at $m=2$ and $m=4$ to obtain the stated formulas.

\subsubsection{Proof of Theorem~\ref{thm:identification}}

\noindent\textbf{Step 1 (Choose a nonzero invariant contrast).}
Since $r=n-q>0$, choose $b\ne0$ in $\mathbb R^r$ and set $a=Rb$.
Then $a\ne0$, $a'D=0$, and $a'Y=b'R'Y$. Equality of the
conditional laws of $R'Y$ therefore implies equality of the
conditional laws of $a'Y$. Under configuration $k$,
\begin{equation}
\label{eq:contrast_model_proof}
a'Y=a'f_{kX}+a'(V_k-U_k).
\end{equation}
The first term is nonrandom given $X$ and affects no cumulant of
order at least two.

\medskip
\noindent\textbf{Step 2 (Identify the half-normal squared scale).}
By Lemma~\ref{lem:contrast_cumulants},
\[
\kappa_4(a'Y\mid X;\vartheta_k)
=\frac{8(\pi-3)}{\pi^2}(\sigma_{u,k}^2)^2\sum_j a_j^4.
\]
The coefficient is positive because $\pi>3$ and $a\ne0$.
Equality of the two fourth cumulants therefore implies
$(\sigma_{u,1}^2)^2=(\sigma_{u,2}^2)^2$. Both squared scales are
nonnegative, so
\begin{equation}
\label{eq:sigu_equal_proof}
\sigma_{u,1}^2=\sigma_{u,2}^2.
\end{equation}
The identified value is explicitly
\[
\sigma_u^2=
\left\{\frac{\pi^2\kappa_4(a'Y\mid X)}
 {8(\pi-3)\sum_j a_j^4}\right\}^{1/2}.
\]
This argument includes the zero-inefficiency boundary.

\medskip
\noindent\textbf{Step 3 (Identify the noise variance).}
Equality of second cumulants and \eqref{eq:sigu_equal_proof} give
\[
\sigma_{v,1}^2\sum_j a_j^2=\sigma_{v,2}^2\sum_j a_j^2.
\]
Since $\sum_j a_j^2>0$,
\begin{equation}
\label{eq:sigv_equal_proof}
\sigma_{v,1}^2=\sigma_{v,2}^2
=\frac{\kappa_2(a'Y\mid X)}{\sum_j a_j^2}
 -(1-2/\pi)\sigma_u^2.
\end{equation}

\medskip
\noindent\textbf{Step 4 (Identify the projected frontier).}
For either configuration,
\[
E(R'Y\mid X;\vartheta_k)
=R'f_{kX}-\sigma_{u,k}\sqrt{2/\pi}\,R'\mathbf1_n.
\]
The nonnegative scales $\sigma_{u,k}$ are equal by Step~2. Equality
of invariant conditional means therefore gives
$R'(f_{1X}-f_{2X})=0$ almost surely. For $h=f_1-f_2$, it follows
that $M_Dh_X=RR'h_X=0$ almost surely.

\medskip
\noindent\textbf{Step 5 (Apply the common-function identification restriction).}
Since $h\in\mathcal F-\mathcal F$, Assumption~1 gives
\begin{equation}
\label{eq:f_equal_proof}
f_1=f_2\quad\mu_X\text{-almost everywhere}.
\end{equation}
This is equivalent to $f_{1X}=f_{2X}$ almost surely: a zero set under
the mixture has zero probability under every cell law, and there
are finitely many cells. Steps~2--5 prove part~1.

\medskip
\noindent\textbf{Step 6 (Use the full law to identify fixed effects).}
Equality of the full conditional laws implies equality of their
projected laws, so part~1 applies. In addition, the full conditional
means satisfy
\begin{equation}
\label{eq:conditional_mean_proof}
E(Y\mid X;\vartheta_k)
=D\delta_k+f_{kX}-\sigma_{u,k}\sqrt{2/\pi}\,\mathbf1_n.
\end{equation}
Canceling the identified half-normal means first gives
\begin{equation}
\label{eq:systematic_equal_proof}
D\delta_1+f_{1X}=D\delta_2+f_{2X}.
\end{equation}
Equivalently,
\begin{equation}
\label{eq:h_in_D_proof}
h_X=D(\delta_2-\delta_1).
\end{equation}
The identified frontier then gives $D(\delta_1-\delta_2)=0$.
Full column rank of $D$ implies
$\delta_1=\delta_2$. The invariant law itself contains no $\delta$,
so this final assertion uses the full law, as claimed.


\subsection{Proof of Corollary~\ref{cor:derivative_identification}}
\label{app:proof_derivative_identification}

\noindent\textbf{Step 1 (Upgrade almost-everywhere equality to pointwise equality).}
Let $h=f_1-f_2$ for two observationally equivalent configurations.
Theorem~\ref{thm:identification} gives $h=0$ $\mu_X$-almost
everywhere. If $h(x_0)\ne0$ at $x_0\in\mathcal X_0$, continuity
gives an open ball contained in $\mathcal X_0$ on which $h$ is
bounded away from zero. Since $x_0\in\operatorname{supp}(\mu_X)$,
that ball has positive $\mu_X$ probability, a contradiction. Thus
$h=0$ everywhere on $\mathcal X_0$.

\medskip
\noindent\textbf{Step 2 (Differentiate inside the open set).}
For $x\in\mathcal X_0$ and a coordinate vector $e_k$, the point
$x+te_k$ remains in $\mathcal X_0$ for all sufficiently small $t$.
Therefore
\[
\frac{h(x+te_k)-h(x)}{t}=0,
\]
and taking the limit gives $\partial h(x)/\partial x_k=0$.
This holds for every coordinate, so $\nabla f_1(x)=\nabla f_2(x)$.
If $x_k=\log z_k$ for physical input $z_k$, the chain rule gives
$\partial\log Y^{\mathrm{frontier}}/\partial\log z_k
=\partial f/\partial x_k$. Summing these identified derivatives
over log inputs gives returns to scale.

\subsection{Proof of Proposition~\ref{prop:gp_completion}}
\label{app:proof_gp_completion}

\noindent\textbf{Step 1 (Project the Gaussian vector).}
Write $K=K_\theta$ and $K_R=R'KR$. Linear transformation of
$f_X\sim N(0,K)$ gives $\eta=R'f_X\sim N(0,K_R)$, and
$\operatorname{Cov}(f_X,\eta)=KR$.
If $K$ is positive definite, then for $b\ne0$,
$b'K_Rb=(Rb)'K(Rb)>0$, because $R$ has full column rank.

\medskip
\noindent\textbf{Step 2 (Separate the part independent of the projection).}
Define $B=KRK_R^{-1}$ and $\xi=f_X-B\eta$. Then
\[
\operatorname{Cov}(\xi,\eta)=KR-BK_R=0,
\qquad
\operatorname{Var}(\xi)=K-KRK_R^{-1}R'K=V_f.
\]
The pair $(\xi,\eta)$ is jointly Gaussian, including when its
covariance is singular. Zero cross-covariance therefore gives
independence, as also follows by factorizing its Gaussian
characteristic function. Conditional on $\eta$, $f_X=B\eta+\xi$,
which proves \eqref{eq:f_given_eta}.

\medskip
\noindent\textbf{Step 3 (Verify support, positivity, and rank).}
Since $R'B=I_r$, $R'm_f=\eta$ and $R'\xi=0$ almost surely.
Thus $R'V_f=0$. As a covariance matrix, $V_f$ is positive
semidefinite. If $K$ is positive definite, set
\[
H_K=K^{1/2}R(R'KR)^{-1}R'K^{1/2}.
\]
This is a symmetric idempotent matrix of rank $r$, and
$V_f=K^{1/2}(I_n-H_K)K^{1/2}$. Multiplication by the invertible
matrix $K^{1/2}$ preserves rank, so
$\operatorname{rank}(V_f)=n-r=q$.

\medskip
\noindent\textbf{Step 4 (Condition using the eliminated likelihood).}
For fixed $(u,\sigma_v^2)$, \eqref{eq:conc_eta} depends on $f_X$
only through $\eta$. Thus, at fixed $(\eta,\theta,u,\sigma_v^2,X)$,
the likelihood factor is constant as a function of the remaining
frontier coordinates. Bayes' formula leaves their conditional GP
law unchanged. This proves the final assertion; it refers to the
eliminated model, rather than to a model with a separate proper
prior supplying additional information on the fixed effects.

\subsection{Proof of Theorem~\ref{thm:score_adjustment}}
\label{app:proof_score_adjustment}

\subsubsection{Conditional Gaussian elimination}
\label{app:conditional_elimination}

To establish \eqref{eq:conditional_integrated_likelihood}, define
$\widehat\delta=(D'D)^{-1}D'e$.
Orthogonality gives
\[
e-D\delta=M_De+D(\widehat\delta-\delta),
\qquad
\|e-D\delta\|^2=e'M_De+
(\delta-\widehat\delta)'D'D(\delta-\widehat\delta).
\]
Also $e'M_De=\|R'e\|^2=\|R'(y+u)-\eta\|^2$. Integrating the last
quadratic term yields
\[
\int_{\mathbb R^q}
\exp\left\{-\frac{(\delta-\widehat\delta)'D'D
                         (\delta-\widehat\delta)}{2\sigma_v^2}\right\}
\,d\delta
=(2\pi\sigma_v^2)^{q/2}|D'D|^{-1/2}.
\]
Multiplication by the original normalizing factor proves
\eqref{eq:conditional_integrated_likelihood}. Profiling sets that
quadratic term to zero and performs no such integration, so it
retains $(2\pi\sigma_v^2)^{-n/2}$.

\subsubsection{Matrix identities for covariance-score calculations}
\label{app:score_matrix_identities}

For the following two lemmas, write $\Sigma=\Sigma_\omega$,
$P=P_\omega$, $\Omega=R'\Sigma R$, and
\begin{align}
A&=D'\Sigma^{-1}D,
\label{eq:Aomega_appendix}\\
P&=\Sigma^{-1}-\Sigma^{-1}DA^{-1}D'\Sigma^{-1}.
\label{eq:Pomega_appendix}
\end{align}
Every inverse exists because $\Sigma$ is positive definite and
$D,R$ have full column rank.

\begin{lemma}[Residual precision and determinant identities]
\label{lem:P_properties_appendix}
The symmetric matrix $P$ satisfies
\begin{align}
&PD=0,\qquad D'P=0,\qquad P\Sigma P=P,
\label{eq:P_sigma_P_appendix}\\
&P=R\Omega^{-1}R',
\label{eq:P_R_appendix}\\
&|\Sigma|\,|D'\Sigma^{-1}D|=|D'D|\,|\Omega|.
\label{eq:det_identity_appendix}
\end{align}
\end{lemma}

\noindent{\em Proof}. 
\noindent\textbf{Step 1 (Verify annihilation and the residual identity).}
Symmetry follows from the definition of $P$. Moreover,
\[
PD=\Sigma^{-1}D-\Sigma^{-1}DA^{-1}A=0,
\]
and symmetry gives $D'P=0$. Multiplication by $\Sigma$ gives
$P\Sigma=I_n-\Sigma^{-1}DA^{-1}D'$. Multiplying on the right
by $P$ then yields $P\Sigma P=P$.

\noindent{\bf Step 2 (Represent $P$ on the contrast subspace).}
Since $P$ is symmetric and annihilates $D$, its range is contained
in $\mathcal C(R)$ and it can be written $P=RBR'$ for a symmetric
$r\times r$ matrix $B$. In addition,
\[
P\Sigma R=(I_n-\Sigma^{-1}DA^{-1}D')R=R.
\]
Premultiplying by $R'$ gives $B\Omega=I_r$, so
$B=\Omega^{-1}$. This proves \eqref{eq:P_R_appendix}.

\medskip
\noindent\textbf{Step 3 (Use an orthogonal block decomposition).}
Let $U_D=D(D'D)^{-1/2}$. Then $G=[U_D,R]$ is an orthogonal
$n\times n$ matrix. The Schur complement of the lower-right block
in $G'\Sigma G$ is
\[
S=U_D'\Sigma U_D-U_D'\Sigma R\Omega^{-1}R'\Sigma U_D.
\]
Block Gaussian elimination gives $|\Sigma|=|\Omega||S|$.
The upper-left block of $(G'\Sigma G)^{-1}$ is both $S^{-1}$
and $U_D'\Sigma^{-1}U_D$. Hence
\[
S^{-1}=(D'D)^{-1/2}A(D'D)^{-1/2},
\qquad |S|^{-1}=|A|/|D'D|.
\]
Substituting in the determinant factorization gives
\eqref{eq:det_identity_appendix}.

\begin{lemma}[Differential of residual precision]
\label{lem:dP_appendix}
For a symmetric perturbation $d\Sigma$ with $D$ fixed,
\begin{align}
\label{eq:dP_appendix}
&dP=-P(d\Sigma)P,\\
&\frac{\partial P_\omega}{\partial\omega_j}
=-P_\omega\Sigma_jP_\omega.
\label{eq:P_derivative_appendix}
\end{align}
\end{lemma}

\noindent{\em Proof}. 
\noindent\textbf{Step 1 (Fix the contrast basis).}
Since $D$ is fixed, choose the same orthonormal basis $R$ of
$\mathcal C(D)^\perp$ for all covariance matrices in a neighborhood
of $\Sigma$. Let $m=\lambda_{\min}(\Sigma)>0$ and take a symmetric
perturbation $H$ with $\|H\|_2<m/2$, where $\|\cdot\|_2$ denotes
the spectral norm. Then $\Sigma+H\succeq(m-\|H\|_2)I_n$ is positive
definite. Write $\Omega_H=R'(\Sigma+H)R=\Omega+R'HR$ and let $P_H$
be the residual precision at $\Sigma+H$. Lemma~\ref{lem:P_properties_appendix}
gives $P=R\Omega^{-1}R'$ and $P_H=R\Omega_H^{-1}R'$.

\medskip
\noindent\textbf{Step 2 (Compute the exact change in residual precision).}
Subtract the inverses of the two contrast covariance matrices using
\[
\Omega_H^{-1}-\Omega^{-1}
=\Omega^{-1}(\Omega-\Omega_H)\Omega_H^{-1}
=-\Omega^{-1}(R'HR)\Omega_H^{-1}.
\]
Multiplying on the left by $R$ and on the right by $R'$ gives
$P_H-P=-P H P_H$. Adding $P H P$ to both sides and substituting
$P-P_H=P H P_H$ on the right yields
\[
P_H-P+P H P=P H(P-P_H)=P H P H P_H.
\]

\noindent\textbf{Step 3 (Bound the remainder and identify the differential).}
Since $R'R=I_r$, the lower bound in Step~1 gives
$\Omega_H\succeq(m/2)I_r$ and hence $\|P_H\|_2\le 2/m$.
The exact remainder in Step~2 therefore satisfies
\[
\|P_H-P+P H P\|_2
\le \|P\|_2^2\,\|H\|_2^2\,\|P_H\|_2
\le \frac{2}{m}\|P\|_2^2\,\|H\|_2^2.
\]
The remainder is therefore $o(\|H\|_2)$ as $H\to0$. Since the map
$H\mapsto-P H P$ is linear, it is the differential of the residual
precision, giving \eqref{eq:dP_appendix}.

\noindent\textbf{Step 4 (Differentiate with respect to a covariance parameter).}
Let $e_j$ be the $j$th coordinate vector and write
$\Sigma_j=\partial\Sigma_\omega/\partial\omega_j$. Varying only
$\omega_j$ gives $H_t=\Sigma_{\omega+t e_j}-\Sigma_\omega
=t\Sigma_j+o(|t|)$. Since $\|H_t\|_2=O(|t|)$, the remainder in
Step~3 divided by $t$ tends to zero. Consequently,
\[
\lim_{t\to0}\frac{P_{\omega+t e_j}-P_\omega}{t}
=-P_\omega\left(\lim_{t\to0}\frac{H_t}{t}\right)P_\omega
=-P_\omega\Sigma_jP_\omega,
\]
which proves \eqref{eq:P_derivative_appendix}.

\subsubsection{Score adjustment, information, and local identification}
\label{app:proof_score_adjustment_steps}

Write $z=z(u)$, $\Sigma=\Sigma_\omega$, $A=A_\omega$,
$P=P_\omega$, and $s=R'z$. The assumed model gives
$s\mid u,X\sim N(0,\Omega)$, where $\Omega=R'\Sigma R$.

\medskip
\noindent\textbf{Step 1 (Integrate the nuisance coefficients exactly).}
The full Gaussian density conditional on $u$ is
\begin{equation}
\label{eq:full_gaussian_appendix}
L(\delta,\omega;y,u)
=(2\pi)^{-n/2}|\Sigma|^{-1/2}
\exp\{-(z-D\delta)'\Sigma^{-1}(z-D\delta)/2\}.
\end{equation}
Its maximizer in $\delta$ is
\begin{equation}
\label{eq:delta_gls_appendix}
\widehat\delta=A^{-1}D'\Sigma^{-1}z.
\end{equation}
Expanding both sides
verifies
\begin{equation}
\label{eq:gls_square_appendix}
(z-D\delta)'\Sigma^{-1}(z-D\delta)
=z'Pz+(\delta-\widehat\delta)'A(\delta-\widehat\delta).
\end{equation}
The Gaussian integral in $\delta$ is $(2\pi)^{q/2}|A|^{-1/2}$.
Therefore
\begin{equation}
\label{eq:LE_appendix}
\int L(\delta,\omega;y,u)\,d\delta
=(2\pi)^{-r/2}|\Sigma|^{-1/2}|A|^{-1/2}e^{-z'Pz/2}.
\end{equation}
Lemma~\ref{lem:P_properties_appendix} gives
$P=R\Omega^{-1}R'$ and $|\Sigma||A|=|D'D||\Omega|$.
Thus this integral is $|D'D|^{-1/2}$ times the normalized
$N(0,\Omega)$ density of $s$. This proves part~1, including both
normalizations in \eqref{eq:general_eliminated_likelihood}.

\medskip
\noindent\textbf{Step 2 (Differentiate the eliminated log likelihood).}
Terms independent of $\omega$ can be suppressed, giving
\begin{equation}
\label{eq:ellE_appendix}
\ell_E=-\tfrac12\log|\Sigma|-\tfrac12\log|A|-\tfrac12z'Pz
+\text{constant}.
\end{equation}
The inverse derivative identity gives
$dA=-D'\Sigma^{-1}(d\Sigma)\Sigma^{-1}D$. Hence
\begin{align*}
d\log|\Sigma|+d\log|A|
&=\operatorname{tr}\{\Sigma^{-1}d\Sigma\}
 -\operatorname{tr}\{A^{-1}D'\Sigma^{-1}(d\Sigma)\Sigma^{-1}D\}\\
&=\operatorname{tr}\{P(d\Sigma)\}.
\end{align*}
Using Lemma~\ref{lem:dP_appendix},
\[
d\ell_E=-\tfrac12\operatorname{tr}\{P(d\Sigma)\}
          +\tfrac12z'P(d\Sigma)Pz.
\]
Set $d\Sigma=\Sigma_j\,d\omega_j$ to obtain
\begin{equation}
\label{eq:score_E_appendix}
S_{E,j}=\tfrac12z'P\Sigma_jPz-\tfrac12\operatorname{tr}(P\Sigma_j),
\end{equation}
which is \eqref{eq:eliminated_score}.

\medskip
\noindent\textbf{Step 3 (Compare with profiling).}
Substitution of $\widehat\delta$ into the full density instead gives
\[
\ell_P=-\tfrac12\log|\Sigma|-\tfrac12z'Pz+\text{constant}.
\]
Differentiation yields
\begin{equation}
\label{eq:score_P_appendix}
S_{P,j}=\tfrac12z'P\Sigma_jPz
          -\tfrac12\operatorname{tr}(\Sigma^{-1}\Sigma_j).
\end{equation}
Subtract this expression from the eliminated score. The difference
is $\tfrac12\operatorname{tr}\{(\Sigma^{-1}-P)\Sigma_j\}$.
Substituting
$\Sigma^{-1}-P=\Sigma^{-1}DA^{-1}D'\Sigma^{-1}$ and cycling
the trace gives
\begin{equation}
\label{eq:adjustment_appendix}
S_{E,j}-S_{P,j}
=\tfrac12\operatorname{tr}
\{A^{-1}D'\Sigma^{-1}\Sigma_j\Sigma^{-1}D\},
\end{equation}
which is \eqref{eq:score_adjustment_term}. This proves part~2.

\medskip
\noindent\textbf{Step 4 (Establish exact score centering).}
Under the generating Gaussian model,
$E(z\mid u,X)=D\delta$ and $\operatorname{Var}(z\mid u,X)=\Sigma$.
For a symmetric matrix $B$, expansion around the mean gives
$E(z'Bz)=\operatorname{tr}(B\Sigma)+(D\delta)'B(D\delta)$.
For $B=P\Sigma_jP$, the mean term vanishes because $PD=0$.
Consequently,
\[
E(z'P\Sigma_jPz\mid u,X)
=\operatorname{tr}(P\Sigma_jP\Sigma)
=\operatorname{tr}(P\Sigma_j),
\]
where the last equality uses $P\Sigma P=P$ and cyclic invariance
of trace. Substitution proves \eqref{eq:score_centering}.
The adjustment in Step~3 is nonrandom conditional on $X$,
so \eqref{eq:profile_score_bias} follows immediately.

\medskip
\noindent\textbf{Step 5 (Derive the expected information).}
Let $g=\Omega^{-1/2}s\sim N(0,I_r)$ and define the symmetric
matrix $B_j=\Omega^{-1/2}\Omega_j\Omega^{-1/2}$, where
$\Omega_j=R'\Sigma_jR$. The score can be written
\[
S_{E,j}=\tfrac12\{g'B_jg-\operatorname{tr}(B_j)\}.
\]
For independent standard normal coordinates,
\[
E(g_ag_bg_cg_d)
=\mathbf1\{a=b\}\mathbf1\{c=d\}
 +\mathbf1\{a=c\}\mathbf1\{b=d\}
 +\mathbf1\{a=d\}\mathbf1\{b=c\}.
\]
This follows by independence when an index appears an odd number
of times, by $E(g_a^2)=1$ for two distinct pairs, and by
$E(g_a^4)=3$ when all indices coincide. Expanding the two
quadratic forms therefore gives
\[
\operatorname{Cov}(g'B_jg,g'B_kg)=2\operatorname{tr}(B_jB_k).
\]
It follows that
\begin{align}
\operatorname{Cov}(S_{E,j},S_{E,k}\mid u,X)
=\tfrac12\operatorname{tr}
 (\Omega^{-1}\Omega_j\Omega^{-1}\Omega_k)=\tfrac12\operatorname{tr}(P\Sigma_jP\Sigma_k).
\label{eq:information_appendix}
\end{align}

To check the expected negative Hessian directly, let
$\Omega_{jk}=\partial^2\Omega/(\partial\omega_j\partial\omega_k)$.
Differentiating the score in contrast coordinates gives
\begin{align*}
\partial_k S_{E,j}
={}&\tfrac12s'\{
-\Omega^{-1}\Omega_k\Omega^{-1}\Omega_j\Omega^{-1}
+\Omega^{-1}\Omega_{jk}\Omega^{-1}
-\Omega^{-1}\Omega_j\Omega^{-1}\Omega_k\Omega^{-1}\}s\\
&+\tfrac12\operatorname{tr}(\Omega^{-1}\Omega_k\Omega^{-1}\Omega_j)
-\tfrac12\operatorname{tr}(\Omega^{-1}\Omega_{jk}).
\end{align*}
Since $E(ss')=\Omega$, the terms containing $\Omega_{jk}$ cancel
in expectation, and the remaining terms give
$-E(\partial_k S_{E,j})
=\tfrac12\operatorname{tr}(\Omega^{-1}\Omega_j\Omega^{-1}\Omega_k)$.
This agrees with \eqref{eq:information_appendix} and proves
the information assertion in part~3.

\medskip
\noindent\textbf{Step 6 (Establish invariance to the contrast basis).}
If $\widetilde R$ is another such basis, then
$\widetilde R=RH$ for an orthogonal $r\times r$ matrix $H$.
Thus $\widetilde s=H's$ and $\widetilde\Omega=H'\Omega H$.
Their determinants and quadratic forms satisfy
\[
|\widetilde\Omega|=|\Omega|,\qquad
\widetilde s'\widetilde\Omega^{-1}\widetilde s
=s'\Omega^{-1}s.
\]
The normal densities coincide because $|H|=1$ in absolute value.
Their parameter derivatives and information matrices coincide as
well. This proves part~4.

\medskip
\noindent\textbf{Step 7 (Characterise positive information).}
Let $d$ be the number of covariance parameters ($d=p+2$ for the
ARD model) and, for $a\in\mathbb R^d$, set
$B(a)=\sum_{j=1}^d a_j\Omega_j$. Then
\begin{equation}
\label{eq:info_frobenius_appendix}
a'\mathcal I_Ea
=\tfrac12\operatorname{tr}\{\Omega^{-1}B(a)\Omega^{-1}B(a)\}
=\tfrac12\|\Omega^{-1/2}B(a)\Omega^{-1/2}\|_F^2.
\end{equation}
Because $\Omega^{-1/2}$ is invertible, this expression is zero
exactly when $B(a)=0$. Hence information is positive definite
if and only if the matrices $\Omega_j$ are linearly independent.

\medskip
\noindent\textbf{Step 8 (Deduce local identification).}
At an interior point satisfying the rank condition, the derivative
of $\omega\mapsto\operatorname{vech}(\Omega_\omega)$ has full
column rank $d$. Select $d$ output coordinates yielding a nonsingular
$d\times d$ Jacobian. The inverse function theorem makes this
selected-coordinate map injective in a neighborhood of that point.
Equality of Gaussian contrast laws implies equality of their
covariance matrices, hence equality of those coordinates and then
of $\omega$ in that neighborhood. This proves the sufficient
local identification claim in part~5. It is not a necessity claim
for identification at singular parameterizations and is not a
global identification argument.


\subsection{Proof of Corollary~\ref{cor:ard_bandwidth_scores}}
\label{app:proof_ard_corollary}

\noindent\textbf{Step 1 (Differentiate the two variance components).}
For stacked indices $a,b$,
\begin{equation}
\label{eq:K_ab_ard_appendix}
[K_\theta]_{ab}=\sigma_f^2
 \exp\left\{-\tfrac12\sum_{k=1}^p[\Delta_k]_{ab}\ell_k^{-2}\right\}.
\end{equation}
Since $\Sigma_\omega=K_\theta+\sigma_v^2I_n$,
$\partial\Sigma/\partial\sigma_v^2=I_n$. Linearity of $K_\theta$
in $\sigma_f^2$ gives
$\partial\Sigma/\partial\sigma_f^2=K_\theta/\sigma_f^2$.

\medskip
\noindent\textbf{Step 2 (Differentiate each log length scale).}
Because $\ell_k=e^{\rho_k}$,
$\partial\ell_k^{-2}/\partial\rho_k=-2\ell_k^{-2}$.
The derivative of the $k$th term of the exponent is therefore
$[\Delta_k]_{ab}/\ell_k^2$. The chain rule yields
\[
\frac{\partial[K_\theta]_{ab}}{\partial\rho_k}
=[K_\theta]_{ab}\frac{[\Delta_k]_{ab}}{\ell_k^2}.
\]
Collecting entries gives
\begin{equation}
\label{eq:Sigma_rho_proof_appendix}
\Sigma_{\rho_k}=K_\theta\odot(\Delta_k/\ell_k^2),
\end{equation}
the Hadamard formula in \eqref{eq:Sigma_sigv_ard}.

\medskip
\noindent\textbf{Step 3 (Substitute into the general score and information).}
Insert this derivative into \eqref{eq:eliminated_score} to obtain
\eqref{eq:ard_score}, and into \eqref{eq:expected_information} to
obtain \eqref{eq:ard_information}. Exact centering follows from
Step~4 of the covariance-score theorem under its stated Gaussian
generating law.

\medskip
\noindent\textbf{Step 4 (State the joint rank requirement precisely).}
Define $G_v=I_r$, $G_f=R'K_\theta R/\sigma_f^2$, and
$G_k=R'\{K_\theta\odot(\Delta_k/\ell_k^2)\}R$.
If $G_v,G_f$ are independent and
\[
\sum_{k=1}^p a_kG_k\in\operatorname{span}\{G_v,G_f\}
\quad\Longrightarrow\quad a_1=\cdots=a_p=0,
\]
then all $p+2$ matrices are independent. Indeed, in any vanishing
linear combination the displayed implication first sets all
length-scale coefficients to zero, and independence of $G_v,G_f$
then sets the other two to zero. Conversely, independence of all
the matrices implies both of these conditions. The final part of
Theorem~\ref{thm:score_adjustment} therefore gives the claimed
regular local identification.

\medskip
\noindent\textbf{Step 5 (Check the alternative parameterization).}
If $\lambda_k=\ell_k^{-2}$ is the parameter, the exponent is
linear in $\lambda_k$ with derivative $-[\Delta_k]_{ab}/2$.
Thus $\partial K_\theta/\partial\lambda_k
=-K_\theta\odot\Delta_k/2$, and all score identities follow
by the chain rule.

\subsection{Proof of Corollary~\ref{cor:conditional_scores}}
\label{app:proof_conditional_scores}

\noindent\textbf{Step 1 (Write the conditional Gaussian experiment).}
Given $(u,\eta,X)$,
\[
R'(y+u)-\eta=R'v\sim N(0,\sigma_v^2I_r).
\]
Let $H$ be its squared norm. The conditional eliminated log
likelihood is
$\ell_E=-(r/2)\log(2\pi\sigma_v^2)-H/(2\sigma_v^2)$.
Differentiating with respect to the squared scale $\sigma_v^2$
gives \eqref{eq:sigma_v_score_E}.

\medskip
\noindent\textbf{Step 2 (Compute the profile adjustment and centering).}
For any completion $f_X$ with $R'f_X=\eta$, profiling over
$\delta$ gives the same $H$ but replaces $r$ by $n$ in the
normalizing term. Thus
\[
S_P=-\frac{n}{2\sigma_v^2}+\frac{H}{2(\sigma_v^2)^2},
\qquad S_E-S_P=\frac{q}{2\sigma_v^2}.
\]
The standardized coordinates of $R'v$ are independent standard
normal variables. Their squared sum gives
$H/\sigma_v^2\sim\chi_r^2$ and hence $E(H)=r\sigma_v^2$.
Substitution yields the two expected scores.

\medskip
\noindent\textbf{Step 3 (Derive the inverse-gamma conditional distribution).}
Multiply \eqref{eq:conc_eta} by the independent prior
$(\sigma_v^2)^{-a_v-1}\exp(-b_v/\sigma_v^2)$. The terms involving
$\sigma_v^2$ become
\[
(\sigma_v^2)^{-(a_v+r/2)-1}
\exp\{-(b_v+H/2)/\sigma_v^2\},
\]
which is the inverse-gamma density in
\eqref{eq:sigma_v_full_conditional}.

\medskip
\noindent\textbf{Step 4 (Verify the conditional oracle comparison).}
Setting the two scores to zero gives $H/r$ and $H/n$, respectively
(with $H>0$ almost surely). Their exact expectations are
$\sigma_v^2$ and $(r/n)\sigma_v^2$. Moreover,
\[
\operatorname{Var}\left(\frac{H}{r\sigma_v^2}\right)=\frac2r.
\]
For fixed $T\geq2$ and $N\to\infty$, $r\to\infty$, so
Chebyshev's inequality gives $H/r\to_p\sigma_v^2$.
Since
\[
\frac{r}{n}=1-\frac1T-\frac1N+\frac{2}{NT}
\longrightarrow\frac{T-1}{T},
\]
we also obtain $H/n\to_p\{(T-1)/T\}\sigma_v^2$.
These calculations supply $(u,\eta)$ at the generating values.
They do not cover additional estimation error in those quantities.

\subsection{Proof of Proposition~\ref{prop:observed_score}}
\label{app:proof_observed_score}

\noindent\textbf{Step 1 (Identify the proper complete-data density).}
At fixed $X$ and $\psi=(\omega',s_u)'$, define
\[
p_C(w,u;\psi)=L_E(\omega;y,u)p(u\mid s_u),
\qquad w=R'y.
\]
For each $u$, $L_E$ is the density of $W$ with mean $-R'u$ and
covariance $\Omega_\omega$. It integrates to one over
$w\in\mathbb R^r$. Also $p(u\mid s_u)$ integrates to one over
$\mathbb R_+^n$. Thus $p_C$ is a proper joint density and
$L_O(\psi;w,X)=\int p_C(w,u;\psi)\,du$ is its marginal density.
This normalization justifies the likelihood score identities without
placing a probability density on the eliminated fixed effects.

\noindent\textbf{Step 2 (Include the half-normal scale score).}
From \eqref{eq:u_joint}, for $s_u>0$ and $u\in\mathbb R_+^n$,
\[
\log p(u\mid s_u)=\frac n2\log(2/\pi)-\frac n2\log s_u
                  -\frac{u'u}{2s_u}.
\]
Differentiation gives $-n/(2s_u)+u'u/(2s_u^2)$.
The other components of $\nabla_\psi\log p_C$ are exactly
$S_E$, proving \eqref{eq:complete_score}.

\noindent\textbf{Step 3 (Differentiate the marginal density).}
The assumed interchange of differentiation and integration gives
\[
\nabla_\psi L_O(w)=\int S_C(w,u)\,p_C(w,u)\,du.
\]
Divide by the positive marginal density $L_O(w)$ and use
$p(u\mid w,X;\psi)=p_C(w,u;\psi)/L_O(w)$ to obtain
\eqref{eq:fisher_identity}. Conditional expectations here are
computed at fixed $\psi$, rather than averaging over a parameter
posterior.

\noindent\textbf{Step 4 (Establish score centering).}
Since $p_C$ integrates to one, differentiation of its integral
gives $E_\psi(S_C\mid X)=0$. For the covariance components,
this also follows by first conditioning on $u$ and applying
Theorem~\ref{thm:score_adjustment}. For the $s_u$ component,
$E(U'U\mid X)=ns_u$, giving the same zero expectation directly.
Iterated expectation in \eqref{eq:fisher_identity} therefore
yields $E_{\psi_0}(S_O\mid X)=0$ under the generating
GP-integrated model.

\noindent\textbf{Step 5 (Differentiate a second time).}
The pointwise density derivative satisfies
\[
\nabla_\psi^2p_C
=p_C\{\nabla_\psi^2\log p_C+S_CS_C'\}.
\]
Integrate this equation over $u$, divide by $L_O(w)$, and subtract
$S_OS_O'$ to convert the second density derivative into the log
density Hessian. This gives
\[
\nabla_\psi^2\log L_O
=E_\psi(\nabla_\psi^2\log p_C\mid w,X)
 +E_\psi(S_CS_C'\mid w,X)-S_OS_O'.
\]
The final two terms are
$\operatorname{Var}_\psi(S_C\mid w,X)$ by Step~3. Multiplying
by minus one proves \eqref{eq:louis_identity}.

\noindent\textbf{Step 6 (Obtain the expected information loss).}
The law of total covariance and $S_O=E(S_C\mid W,X)$ give
\begin{align*}
\operatorname{Var}_\psi(S_C\mid X)
=\operatorname{Var}_\psi(S_O\mid X)+E_\psi\{\operatorname{Var}_\psi(S_C\mid W,X)\mid X\}.
\end{align*}
The assumed differentiability of the normalized densities gives
the usual information equality directly: integration of
$\nabla^2p=p\{\nabla^2\log p+SS'\}$ sets its integral to zero,
so $-E(\nabla^2\log p)=E(SS')$ for both the complete and observed
densities. As their scores have mean zero, these second moments
are the score covariance matrices. Rearranging the total
covariance identity therefore proves
\eqref{eq:missing_information}. A conditional covariance matrix
is positive semidefinite, so the displayed matrix inequality
follows as well.

\noindent\textbf{Step 7 (Explain the boundary restriction).}
At $s_u=0$, the law of $U$ is concentrated at the origin and is
not represented by \eqref{eq:u_joint} with $s_u$ set to zero.
Thus the interior derivatives used in Steps~2--6 cannot be
evaluated at that boundary. This does not affect the separate
population identification argument, which included the point
mass directly.

\section{Prior calibration and computational details}
\label{app:computational_details}

This appendix records implementation details for the posterior in
\eqref{eq:post_joint}. It uses the notation of Section~\ref{sec:bayes},
including $z_u=R'(y+u)$, $K_R=R'K_\theta R$, and
$\Omega_\theta=K_R+s_vI_r$.

\subsection{Prior calibration}
\label{app:prior_calibration}

Prior scales are fixed before posterior simulation. For a nonconstant
input, one possible calibration is $\xi_{\ell,k}=c_{\ell,k}/d_k^2$,
where $d_k$ is the median of its strictly positive pairwise distances
and $c_{\ell,k}>0$ is fixed. A constant input supplies no information
about its length scale and requires separate treatment. If the signal
scale is calibrated using the response, a basis-invariant choice is
\[
\xi_f=c_f\max\left\{\frac{y'M_Dy}{r},\epsilon_y\right\},
\qquad c_f>0,\quad \epsilon_y>0,
\]
where $\epsilon_y$ is a fixed positive floor in squared log-output
units. The ordinary sample variance of the coordinates of $R'y$ is
generally basis dependent. Response-based calibration gives an
empirical Bayes analysis conditional on the selected scale; it should
be reported with prior-sensitivity checks.

\subsection{Posterior propriety}
\label{app:posterior_propriety}

The posterior is proper under these priors. Indeed,
$0<L_E\leq(2\pi s_v)^{-r/2}$; integrating the other proper prior
blocks leaves a finite upper bound because
\[
E(s_v^{-r/2})=b_v^{-r/2}
\frac{\Gamma(a_v+r/2)}{\Gamma(a_v)}<\infty.
\]
Positivity of the Gaussian likelihood also gives a strictly positive
normalizing constant. This establishes propriety for each fixed
choice of the positive prior scales, including a fixed empirical
Bayes calibration.

\subsection{Simulation of the conditional posterior blocks}
\label{app:conditional_simulation}

An exact conditional simulation avoids forming the covariance
difference: independently draw $\widetilde\eta\sim N(0,K_R)$ and
$\widetilde\varepsilon\sim N(0,s_vI_r)$, and set
\[
\eta=\widetilde\eta+K_R\Omega_\theta^{-1}
      (z_u-\widetilde\eta-\widetilde\varepsilon).
\]
Its mean is $\bar\eta$ and its covariance is $V_\eta$ by direct
expansion.

For the inefficiency update, let $r_\eta=y-R\eta$ as in
Section~\ref{subsec:eta_post}.
For moderate dimensions, joint simulation can use the minimax-tilting
method of \citet{Botev2017}. Alternatively, write $m_{jk}=[M_D]_{jk}$
and update coordinates sequentially using
\begin{equation}
\label{eq:post_uj_cond}
u_j\mid u_{-j},\eta,s_u,s_v,y
\sim TN_{[0,\infty)}(\bar u_j,s_j^2),
\end{equation}
with
\begin{align*}
s_j^2&=(s_u^{-1}+m_{jj}s_v^{-1})^{-1},\\
\bar u_j&=-\frac{s_j^2}{s_v}
\left\{(M_Dr_\eta)_j+\sum_{k\ne j}m_{jk}u_k\right\}.
\end{align*}
Here $TN_{[0,\infty)}(m,s^2)$ denotes an $N(m,s^2)$ variable
conditioned to be nonnegative. It is generally not a half-normal
distribution. One complete coordinate sweep preserves the joint
conditional law in \eqref{eq:post_u_tmvn}; it need not produce an
independent draw from that law.

The balanced-panel projector can be applied without storing a dense
$n\times n$ matrix. For a stacked vector $h$,
\[
(M_Dh)_{it}=h_{it}-\bar h_{i\cdot}-\bar h_{\cdot t}
                 +2\bar h_{\cdot\cdot},
\qquad m_{jj}=1-\frac1N-\frac1T+\frac{2}{NT}.
\]
The coefficient two retains the grand mean under the zero-sum fixed
effects. Maintaining row, column, and total sums permits inexpensive
coordinate updates.

\subsection{Collapsed updates of the GP hyperparameters}
\label{subsec:gp_hyper}

Integrating $\eta$ from its Gaussian block gives
$z_u\mid\theta,s_v,u\sim N(0,\Omega_\theta)$. At fixed $(u,s_v)$,
the collapsed conditional posterior is
\begin{equation*}
\pi_E(\theta\mid u,s_v,y)
\propto |\Omega_\theta|^{-1/2}
\exp\left(-\frac12z_u'\Omega_\theta^{-1}z_u\right)p(\theta).
\end{equation*}
For computation, use unconstrained parameters
\[
\phi=(\psi_f,\rho_1,\ldots,\rho_p)',\qquad
\psi_f=\log s_f,\qquad \rho_k=\log\ell_k.
\]
Let $b_f=(2\xi_f)^{-1}$ and $b_{\ell,k}=(2\xi_{\ell,k})^{-1}$.
The prior Jacobians are explicit: $ds_f/d\psi_f=s_f$ and
$|d\lambda_k/d\rho_k|=2\lambda_k$, where
$\lambda_k=e^{-2\rho_k}$. Thus the log posterior density with
respect to $d\phi$, up to an additive constant, is
\begin{align}
\ell_c(\phi;u,s_v)
=-\frac12\log|\Omega_\theta|
    -\frac12z_u'\Omega_\theta^{-1}z_u
    +\frac12\psi_f-b_fe^{\psi_f}
-\sum_{k=1}^p\left(\rho_k+b_{\ell,k}e^{-2\rho_k}\right).
\label{eq:post_phi_target}
\end{align}
These Jacobians are part of the target density and must be included
in Metropolis acceptance probabilities.

For a symmetric Gaussian random-walk proposal
$\phi^*=\phi+\varepsilon$, $\varepsilon\sim N(0,C_\phi)$, the
acceptance probability is
\[
\alpha_\phi=\min\{1,\exp[\ell_c(\phi^*;u,s_v)
                            -\ell_c(\phi;u,s_v)]\}.
\]
The proposal covariance $C_\phi$ is positive definite and fixed
during the retained sampling phase. The proposal and target in this
step do not condition on the old value of $\eta$.

The covariance-score formulas in Section~\ref{subsec:score_adjustment}
also give the transformed posterior gradient. With
$\Omega_j=\partial\Omega_\theta/\partial\phi_j$,
\begin{equation*}
\frac{\partial\ell_c}{\partial\phi_j}
=\frac12z_u'\Omega_\theta^{-1}\Omega_j\Omega_\theta^{-1}z_u
 -\frac12\operatorname{tr}(\Omega_\theta^{-1}\Omega_j)+b_j(\phi),
\end{equation*}
where
\begin{align*}
\Omega_{\psi_f}&=R'K_\theta R,
&b_{\psi_f}(\phi)&=\frac12-b_fe^{\psi_f},\\
\Omega_{\rho_k}&=R'\{K_\theta\odot(\Delta_k/\ell_k^2)\}R,
&b_{\rho_k}(\phi)&=-1+2b_{\ell,k}e^{-2\rho_k}.
\end{align*}
These expressions provide checks on numerical derivatives or support
gradient-based proposals with the appropriate proposal-density ratio.
The baseline sampler uses the symmetric random walk above.

\subsection{An optional variance-preserving move}
\label{subsec:ridge_move}

Dependence between inefficiency and the two error scales can make
separate updates inefficient. An additional Metropolis move changes
these quantities jointly while preserving the composed-error variance.
Let $\kappa=1-2/\pi$. Keeping $(\eta,\theta)$ fixed, draw
$a\sim N(0,\tau^2)$, set $c=e^a$, and propose
\begin{equation}
\label{eq:post_ridge_proposal}
u^*=cu,\qquad s_u^*=c^2s_u,\qquad
s_v^*=s_v+\kappa(1-c^2)s_u.
\end{equation}
Reject if $s_v^*\leq0$. Otherwise $s_v^*+\kappa s_u^*=s_v+\kappa s_u$,
and the acceptance probability is
\begin{equation}
\label{eq:post_ridge_acceptance}
\alpha_c=\min\left\{1,\ c^2
\left(\frac{s_v^*}{s_v}\right)^{-r/2}
\exp\left(-\frac{H^*}{2s_v^*}+\frac{H}{2s_v}\right)
\frac{p(s_u^*)p(s_v^*)}{p(s_u)p(s_v)}\right\},
\end{equation}
where $H=\|R'(y+u)-\eta\|^2$ and
$H^*=\|R'(y+u^*)-\eta\|^2$, evaluated at the current and proposed
states, respectively.

The factor $c^2$ is required for detailed balance. On the augmented
state, the transformation
$(u,s_u,s_v,a)\mapsto(u^*,s_u^*,s_v^*,-a)$ is its own inverse
and has absolute Jacobian $c^{n+2}$. The proposal density of $a$
is symmetric, while the half-normal density ratio is
$p(cu\mid c^2s_u)/p(u\mid s_u)=c^{-n}$. Multiplying these factors
with the remaining posterior ratio gives
\eqref{eq:post_ridge_acceptance}. The GP prior is unchanged because
$(\eta,\theta)$ are held fixed.

This move preserves the variance but generally changes the mean and
higher cumulants of the composed error; the likelihood ratio remains
essential. It is optional, and its effect on mixing must be assessed
from the chains. The proposal scale $\tau$ and the number of attempts
per iteration are fixed after warm-up.

\section{Posterior estimates of the translog coefficients}
\label{app:translog}

\begin{table}[!htbp]\centering\small
\caption{Posterior estimates of the translog coefficients.}\label{tab:translog_coef}
\begin{tabular}{lrrrr}
\toprule
 & \multicolumn{2}{c}{TWFE--TL} & \multicolumn{2}{c}{FE--TL} \\
\cmidrule(lr){2-3}\cmidrule(lr){4-5}
Coefficient & Mean & 95\% credible interval & Mean & 95\% credible interval \\
\midrule
$\beta_0$ & 0.0183 & ($-$0.0004, 0.0364) & $-$0.1083 & ($-$0.1390, $-$0.0779) \\
$\beta_{\mathrm{Labor}}$ & 0.0071 & ($-$0.0178, 0.0317) & 0.0270 & ($-$0.0240, 0.0783) \\
$\beta_{\mathrm{Cows}}$ & 0.6027 & (0.5519, 0.6536) & 0.6656 & (0.5572, 0.7738) \\
$\beta_{\mathrm{Feed}}$ & 0.3053 & (0.2757, 0.3346) & 0.2592 & (0.1933, 0.3268) \\
$\beta_{\mathrm{Land}}$ & $-$0.0171 & ($-$0.0542, 0.0203) & 0.0199 & ($-$0.0359, 0.0761) \\
$\beta_{\mathrm{Roughage}}$ & 0.0424 & (0.0229, 0.0616) & 0.0217 & ($-$0.0124, 0.0545) \\
\addlinespace
\multicolumn{5}{l}{\textit{Second-order terms}} \\
$\beta_{\mathrm{Labor},\mathrm{Labor}}$ & 0.0748 & ($-$0.0128, 0.1603) & 0.0417 & ($-$0.0542, 0.1395) \\
$\beta_{\mathrm{Cows},\mathrm{Cows}}$ & 0.2769 & (0.0132, 0.5398) & 0.3493 & (0.0267, 0.6691) \\
$\beta_{\mathrm{Feed},\mathrm{Feed}}$ & 0.2011 & (0.1145, 0.2895) & 0.1793 & (0.0919, 0.2694) \\
$\beta_{\mathrm{Land},\mathrm{Land}}$ & 0.0441 & ($-$0.0554, 0.1450) & 0.0143 & ($-$0.0856, 0.1168) \\
$\beta_{\mathrm{Roughage},\mathrm{Roughage}}$ & 0.0116 & ($-$0.0218, 0.0439) & 0.0077 & ($-$0.0256, 0.0411) \\
$\beta_{\mathrm{Labor},\mathrm{Cows}}$ & $-$0.0097 & ($-$0.1059, 0.0867) & $-$0.0327 & ($-$0.1613, 0.0947) \\
$\beta_{\mathrm{Labor},\mathrm{Feed}}$ & $-$0.0612 & ($-$0.1196, $-$0.0028) & $-$0.0270 & ($-$0.1082, 0.0549) \\
$\beta_{\mathrm{Labor},\mathrm{Land}}$ & 0.0463 & (0.0035, 0.0887) & 0.0225 & ($-$0.0351, 0.0788) \\
$\beta_{\mathrm{Labor},\mathrm{Roughage}}$ & $-$0.0020 & ($-$0.0279, 0.0238) & 0.0091 & ($-$0.0250, 0.0437) \\
$\beta_{\mathrm{Cows},\mathrm{Feed}}$ & $-$0.3048 & ($-$0.4240, $-$0.1884) & $-$0.3230 & ($-$0.4539, $-$0.1942) \\
$\beta_{\mathrm{Cows},\mathrm{Land}}$ & $-$0.0965 & ($-$0.2184, 0.0269) & $-$0.0440 & ($-$0.1764, 0.0896) \\
$\beta_{\mathrm{Cows},\mathrm{Roughage}}$ & 0.1337 & (0.0602, 0.2077) & 0.1210 & (0.0415, 0.1976) \\
$\beta_{\mathrm{Feed},\mathrm{Land}}$ & 0.1210 & (0.0518, 0.1900) & 0.1197 & (0.0500, 0.1908) \\
$\beta_{\mathrm{Feed},\mathrm{Roughage}}$ & $-$0.0789 & ($-$0.1217, $-$0.0355) & $-$0.0794 & ($-$0.1235, $-$0.0350) \\
$\beta_{\mathrm{Land},\mathrm{Roughage}}$ & $-$0.0490 & ($-$0.0938, $-$0.0039) & $-$0.0398 & ($-$0.0865, 0.0058) \\
\addlinespace
\multicolumn{5}{l}{\textit{Time trend}} \\
$\beta_t$ & -- & -- & 0.0321 & (0.0245, 0.0395) \\
$\beta_{tt}$ & -- & -- & $-$0.0031 & ($-$0.0042, $-$0.0020) \\
$\beta_{\mathrm{Labor},t}$ & -- & -- & $-$0.0041 & ($-$0.0111, 0.0029) \\
$\beta_{\mathrm{Cows},t}$ & -- & -- & $-$0.0066 & ($-$0.0202, 0.0069) \\
$\beta_{\mathrm{Feed},t}$ & -- & -- & 0.0055 & ($-$0.0029, 0.0135) \\
$\beta_{\mathrm{Land},t}$ & -- & -- & $-$0.0047 & ($-$0.0103, 0.0009) \\
$\beta_{\mathrm{Roughage},t}$ & -- & -- & 0.0029 & ($-$0.0010, 0.0070) \\
\bottomrule
\end{tabular}
\end{table}

\end{document}